\documentclass[aps,prb,twocolumn,superscriptaddress,floatfix,amsmath,onlytext,fullfamily,textlf,longbibliography]{revtex4-2} % standart PRL preprint class
  
\usepackage[normalem]{ulem}
\usepackage{bm} %Access bold symbols in maths mode: The bm package defines a command \bm which makes its argument bold. The argument may be any maths object from a single symbol to an expression.
\usepackage{amssymb}
\usepackage{amsfonts} %TeX fonts from the American Mathematical Society: An extended set of fonts for use in mathematics, including: extra mathematical symbols; blackboard bold letters (uppercase only); fraktur letters; subscript sizes of bold math italic and bold Greek letters; subscript sizes of large symbols such as sum and product; added sizes of the Computer Modern small caps font; cyrillic fonts (from the University of Washington); Euler mathematical fonts.
\usepackage{amsmath} %AMS mathematical facilities for LaTeXg
\usepackage{graphicx} % Include figure files
\usepackage{xcolor} %Driver-independent color extensions for LaTeX and pdfLaTeX: The package starts from the basic facilities of the color package, and provides easy driver-independent access to several kinds of color tints, shades, tones, and mixes of arbitrary colors. Colors can be mixed like \color{red!30!green!40!blue}

\usepackage[pdftex,colorlinks=true,allcolors=blue,breaklinks=true]{hyperref}  % The build LaTeX => PDF
\usepackage[utf8]{inputenc} % to insert the character directly by copy paste or as ^+i typed on your keyboard
\usepackage{textgreek}
\definecolor{g-blue}{rgb}{0.83,0.95,1}
\definecolor{g-yellow}{rgb}{1,1,0.7}
\definecolor{g-green}{rgb}{0.9,1,0.9}
\definecolor{green}{rgb}{0,0.6,0}
\definecolor{cyan}{rgb}{0,0.7,0.7}
\definecolor{black}{rgb}{0,0,0}
\definecolor{grey}{rgb}{0.4,0.4,0.4}
\definecolor{nature-blue}{rgb}{0.0,0.200,0.500}

\def\white#1{\textcolor{white}{#1}}

\def \ed {\end{document}}
\def\Fbox#1{\vskip1ex\hbox to 8.5cm{\hfil\fboxsep0.3cm\fbox{%
		\parbox{8.0cm}{#1}}\hfil}\vskip1ex\noindent}  %%  {TEXT} in BOX

\def\be{\begin{equation}}
\def\ee{\end{equation}}
\def\bea{\begin{eqnarray}}
\def\eea{\end{eqnarray}}
\def\bse{\begin{subequations}}
\def\ese{\end{subequations}}

\let \= \equiv  
\let\*\cdot 
\let\^\widehat 
\let\-\overline

\def\1{\bm1}

\def\<{\left\langle}    \def\>{\right\rangle}
\def\({\left(}          \def\){\right)}
\def\[ {\left[}         \def\]{\right]}

\newcommand{\eq}[1]{(\ref{#1})}%%  requires \eq{label}
\newcommand{\Eq}[1]{Eq.\,(\ref{#1})}%%  requires \eq{label}
\newcommand{\Eqs}[1]{Eqs.\,(\ref{#1})}%%  requires \eq{label}
\newcommand{\Fig}[1]{Fig.\,\ref{#1}}%%  requires \Fef{label}
\newcommand{\Figs}[1]{Figs.\,\ref{#1}}%%  requires \Fef{label}
\newcommand{\Sec}[1]{Sec.\,\ref{#1}}%%  requires \Fef{label}
\newcommand{\Secs}[1]{Secs.\,\ref{#1}}%%  requires \Fef{label}
\newcommand{\REF}[1]{Ref.\,\cite{#1}}%%  requires \Fef{label}
\newcommand{\Refs}[1]{Refs.\,\cite{#1}}%%  requires \Fef{label}
\newcommand{\B}[1]{{\bm{#1}}}%% Bold Roman & Greek Lower & Upper Case
\newcommand{\C}[1]{{\mathcal{#1}}}    %%   Calligrapfic Upper case
\renewcommand{\sb}[1]{_{\text {#1}}}  %% sub-   for lower case
\renewcommand{\sp}[1]{^{\text {#1}}}  %% super- for lower case
\newcommand{\Sp}[1]{^{^{\text {#1}}}} %% Super- for Upper case
\def\Sb#1{_{\scriptscriptstyle\rm{#1}}}
\begin{document}

	\title{ Bose-Einstein condensation in systems with flux equilibrium }
 
	\author{Victor~S.~L'vov}
 	\email{victor.lvov@gmail.com}	
    \affiliation{Department of Physics of Complex Systems, Weizmann Institute of Science, Rehovot 76100, Israel}
	
    \author{Anna Pomyalov}
 	\email{anna.pomyalov@weizmann.ac.il}	
	\affiliation{Department of Chemical and Biological Physics, Weizmann Institute of Science, Rehovot 76100, Israel}
	
    \author{Sergey~V.~Nazarenko}
 	\email{sergey.nazarenko@unice.fr}
	\affiliation{Universit\'{e} C$\hat{o}$te d'Azur, CNRS, Institut de Physique de Nice, 17 rue Julien Laupr$\hat{e}$tre 06200 Nice, France}
	
    \author{Dmytro~A.~Bozhko}
	\email{dbozhko@uccs.edu}
	\affiliation{Department of Physics and Energy Science, University of Colorado Colorado Springs, Colorado Springs, Colorado 80918, USA}

    \author{Alexander~J.~E.~Kreil}
   	\email{kreil@rhrk.uni-kl.de}
	\affiliation{Fachbereich Physik and Landesforschungszentrum OPTIMAS, Rheinland-Pf\"alzische Technische Universit\"at Kaiserslautern-Landau, 67663 Kaiserslautern, Germany}
    
    \author{Burkard~Hillebrands}
	\email{hilleb@rptu.de}
	\affiliation{Fachbereich Physik and Landesforschungszentrum OPTIMAS, Rheinland-Pf\"alzische Technische Universit\"at Kaiserslautern-Landau, 67663 Kaiserslautern, Germany}
	
    \author{Alexander~A.~Serga}
	\email{serha@rptu.de}
	\affiliation{Fachbereich Physik and Landesforschungszentrum OPTIMAS, Rheinland-Pf\"alzische Technische Universit\"at Kaiserslautern-Landau, 67663 Kaiserslautern, Germany}

\begin{abstract}
We consider flux equilibrium in dissipative nonlinear wave systems subject to external energy pumping.
In such systems, the elementary excitations, or quasiparticles, can create a Bose-Einstein condensate.
We develop a theory on the Bose-Einstein condensation of quasiparticles for 
various regimes of external excitation, ranging from weak and stationary to ultra-strong pumping, enabling us to determine the number of quasiparticles near the bottom of the energy spectrum and their distribution along wave vectors.  We identify physical phenomena leading to condensation in each of the regimes.
For weak stationary pumping, where the distribution of quasiparticles deviates only slightly from thermodynamic equilibrium, we define a range of pumping parameters where the condensation occurs and estimate the density of the condensate and the fraction of the condensed quasiparticles.
As the pumping amplitude increases, a powerful influx of injected quasiparticles is created by the Kolmogorov-Zakharov scattering cascade, leading to their Bose-Einstein condensation. With even stronger pumping, kinetic instability may occur, resulting in a direct transfer of injected quasiparticles to the bottom of the spectrum. For the case of ultra-strong parametric pumping, we have  developed 
%created
a stationary nonlinear theory of kinetic instability. The theory agrees qualitatively with experimental data obtained using Brillouin light scattering spectroscopy during parametric pumping of magnons in room-temperature films of yttrium-iron garnet.
\end{abstract}

\maketitle 

%\tableofcontents 

%%%%%%%%%%%%%%%%%%%%%%%%%%%%%%%%%
\section{Introduction}  
The Bose-Einstein (BE) condensate  (BEC) is a state of matter with a macroscopically large number of bosons occupying the lowest quantum state and demonstrating coherence at macroscopic scales \cite{Stone2015, Einstein1925in2005, Bose1924, Froelich1968, Snoke2006}. This phenomenon was observed and investigated in atomic systems such as $^4$He, $^3$He (in the latter, the role of bosons is played by Cooper pairs of fermionic $^3$He atoms), and in ultra-cold trapped atoms \cite{Davis1995, Anderson1995}.

%%%%%%%%%%%%%%%%%%%%
\begin{figure}  
  \includegraphics[width=1\columnwidth]{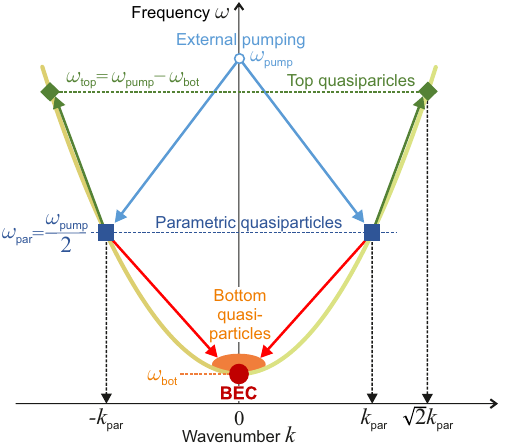} 
        \caption{\label{f:0}
        Schematic representation of the parabolic frequency spectra of quasiparticles \,\eqref{3} with their relevant groups: parametric quasiparticles with frequency $\omega\sb {par}$, denoted as blue squares {\color[rgb]{0.184,0.333,0.592}$\blacksquare$}, BEC   with frequency $\omega_0$, denoted as red circles {\color[rgb]{0.753,0,0}\LARGE \textbullet}, and top quasiparticles with the frequency $\omega\sb{top}=2\omega\sb{par}-\omega\sb{bot}$, denoted as  green diamonds {\color[rgb]{0.329,0.510,0.208}$\LARGE \blacklozenge$}.
        The bottom quasiparticles slightly above the frequency $\omega\sb{bot}$ are shown by the orange area.  The light-blue arrows denote the process of creation of parametric quasiparticles by external pumping. The red and green arrows show the process of four-wave scattering, leading to the phenomenon of kinetic instability.  }
\end{figure}
%%%%%%%%%%%%%%%%%%%%
	
BECs were also found in systems of bosonic quasiparticles such as polaritons \cite{Amo2009} and excitons \cite{Eisenstein2004} in semiconductors, photons in micro-cavities \cite{Klaers2010}, as well as magnons in superfluid $^3$He \cite{Bunkov2008} and magnetic crystalline materials \cite{Demokritov2006, Serga2014a, Schneider2020}. In all these cases quasiparticles have a finite lifetime, and the appearance of 
steady-state BEC requires continuous or periodic excitation  (pumping) of quasiparticles by an external source. To some extent, these systems can be considered as being in a flux-defined (rather than thermodynamic) equilibrium. This feature makes the quasiparticle systems qualitatively different from the systems of real particles (atoms) whose total number is conserved. BECs of quasiparticles are drawing significant interest for their possible
applications in new technologies of information transfer and data processing, including coherent quantum optics (see, e.g., \cite{Turschmann2019,Fukami2021,Lai2007}). The investigation of such flux-driven systems is the main motivation of the present work. 
	
In this paper,  we compare various scenarios of the evolution of a weakly interacting overpopulated gas of quasiparticles toward BEC under conditions of flux equilibrium. For simplicity, we consider an isotropic homogeneous wave system with a  parabolic dispersion law:
\begin{subequations}\label{Ek}
\begin{equation}\label{3}
\omega_k=\omega_0[1 +(ak)^2] \ .
\end{equation}
Here $\omega_0$ is the gap of the frequency spectrum (``bottom'' frequency), $a$ is a characteristic scale, and $k$ is the wave number, see \Fig{f:0}. This choice simplifies the comparison of a BEC of quasiparticles with the energy $E_k=\hbar \omega_k$ in the flux-equilibrium wave systems with a BEC of non-relativistic bosons with  the energy 
\begin{equation}\label{Ek1}
E_k= (\hbar k)^2/(2 m)
\end{equation}
expressed via the Plank's constant $h=2\pi \hbar$ and the particle mass $m$.   Expressions for $E_k$  stress the wave-particle duality in describing waves and quasiparticles in quantum mechanics. For example, using the parabolic dispersion law\,\eqref{3}, we write an equation for energy similar to \Eq{Ek1}, 
 \begin{equation}\label{Ek3}
    E_k =E_0+(\hbar k)^2/(2m )\,, \quad E_0\= \hbar \omega_0\ .
\end{equation}
Thus, we can consider quasiparticles having the energy spectrum $E_k=\hbar \omega_k$ with  $\omega_k$ defined by \Eq{3} and an effective mass 
\begin{equation}\label{mass}
m\sb{eff} = \hbar /(2 \omega_0 a^2) \ .
\end{equation}\end{subequations}

 To simplify the discussion further we assume that the pumping of the system results in the appearance of quasiparticles with a  particular frequency $\omega\sb {par}$.  
A well-known example of such pumping is the parametric excitation of magnons in a ferromagnetic material by an almost homogeneous external electromagnetic field of frequency $\omega\sb{pump}=2\omega\sb {par}$. In this case, parametric magnons with frequency $\omega\sb {par}= \omega(\pm \B k\sb {par})$ appear as a result of the decay process with the conservation law:
	\begin{equation}\label{cl}
	\omega\sb{pump}=\omega( \B k\sb  {par})+ \omega(-\B k\sb  {par})\,,
	\end{equation}
schematically shown by two blue arrows in \Fig{f:0}. All our results can be easily generalized for more sophisticated quasiparticle pumping in a wide range of frequencies, for example by parametric pumping with a noisy electromagnetic field (or noise modulation of quasiparticle frequency) \cite{Zautkin1983}.

The paper aims to investigate the processes  leading to 
the emergence of Bose-Einstein condensates (BECs) in various parameter ranges of the pumping. It also seeks to determine the total number of quasiparticles ($N\sb{tot}$) in the vicinity of the spectral minimum $\omega_0 $ and the fraction  $N\Sb{BEC}$ of the number of quasiparticles that constitute the condensed part.

The structure of the paper is as follows. In \Sec{s:background} we describe the wave system under consideration and, to introduce notations, remind the well-known results for BE condensation of bosons in three-dimensional (3D) and two-dimensional (2D) systems.   
	
Next, in \Sec{s:FE-WP} we formulate criteria for the BE condensation and find $N\sb{tot}$ and $N\Sb{BE}$ for a relatively simple case of weak pumping, for which the nonlinear wave system, even in the presence of the energy and particle number fluxes, is close to the thermodynamic equilibrium. In \Sec{s:FE-SP}
we consider the case of strong pumping. Here, in the 3D system, the overpopulated gas of quasiparticles is transferred by step-by-step cascade processes down the frequency band, followed by the thermalization of low-energy quasiparticles into the BEC state \cite{Kalafati1991,Chumak2009a}. We also discuss more involved 2D and thin-film cases. 

A very strong pumping regime, as considered in \Sec{s:KI}, to the best of our knowledge, is currently realized only for magnons in ferromagnetic materials. However, the physical picture in this regime does not depend on the specific properties of magnons. We, therefore, consider quasiparticles with a generic parabolic frequency spectrum, shown in Fig.\,\ref{f:0}. Here, the cascade process can be accompanied by a direct transfer of the parametrically injected quasiparticles to the lowest (bottom) and high (top) energy states by a $2\Leftrightarrow2$ scattering process \cite{Lavrinenko1981, Melkov1991, Melkov1994}
\begin{eqnarray}\label{KI-1}
\omega (\B k\sb{par})+ \omega (\B k\sb{par}')\Rightarrow\omega (\B k\sb{bot})+\omega(\B k\sb{top})\ .
\end{eqnarray}
 
In this process, referred to as the kinetic instability (KI)\,\cite{Lavrinenko1981}, a dense cloud of incoherent ``bottom'' quasiparticles is formed close to the BEC point. This scattering process is sketched in \Fig{f:0} by red arrows pointing from blue-filled squares to the orange area and discussed in \Sec{ss:LKI}. By the energy conservation law\,\eqref{KI-1}, the same number of parametric quasiparticles is transferred to higher energy states with frequency $\omega \sb{top}\simeq 2\omega\sb{p}- \omega \sb b$  and energy above thermodynamic equilibrium (top quasiparticles, shown in \Fig{f:0} by filled green diamonds). The feedback influence of the top and bottom quasiparticles on the parametric ones is studied in \Sec{ss:S-theory}.
Note that in the considered case of the parabolic isotropic dispersion surface, the momentum conservation law is satisfied  because 
the parametric quasiparticles fill the entire isofrequency circle $\omega_\mathrm{par}$. Thus, the scattering process involves parametric quasiparticles with wave vectors $k_\mathrm{par}$ arranged at an angle of $45^\circ$ to each other, which ensures that the wave vector length of the top quasiparticles is $\sqrt{2}k_\mathrm{par}$ (see \Fig{f:0}). 
Four-particle scattering of parametric and bottom quasiparticles is responsible for the widening of the package of the bottom quasiparticles. These processes are considered in the framework of the nonlinear theory of kinetic instability, developed in \Secs{ss:scat} and \ref{ss:NLT}.  

Section \ref{s:EXP} is devoted to the experimental study of the BE condensation of magnons in thin films of yttrium iron garnet (YIG, $\mathrm{Y}_3\mathrm{Fe}_5\mathrm{O}_{12}$) using Brillouin light scattering (BLS) spectroscopy. 
This ferrimagnetic material is a classical material for the experimental study of nonlinear magnon dynamics. There are several reasons for this: 
(i)  most importantly, it has the lowest known spin-wave damping; 
(ii) its Curie temperature $T_\mathrm{C}=560$\,K allows experiments to be carried out at room temperature; 
(iii) being a dielectric, it is transparent to microwave electromagnetic radiation, which makes it possible to excite magnetic oscillations in the entire volume of bulk samples and study them using common microwave techniques; 
(iv) thin single-crystal films of YIG are transparent to visible light, which enables the study of magnon dynamics also by optical methods with spatial, temporal, frequency, and wave vector resolution. 
In our experiments, magnons were pumped by an external electromagnetic field, as shown in \Fig{f:2}. Comparing the nonlinear theory of kinetic instability with available and new experimental results, we conclude that at large pumping amplitudes, kinetic instability is the main channel for transferring magnons from the pumping region directly to the lower part of their frequency spectrum.   We confirm several predictions of the newly-developed nonlinear theory of BEc of quasiparticles. 
%%%%%%%%%%%%%%%%%%%%%%%%%%%%%%%%%
\section{\label{s:background} BEC in thermodynamic equilibrium: Analytical background}
The physics of a BEC in systems with a flux equilibrium is, in many aspects, similar to that of a BEC in thermodynamic equilibrium. Therefore, to stress the similarities and differences of basic physics in these two  regimes,
%cases
 it is useful to use the same or similar notations. To introduce these notations, we shortly describe the BEC process in wave systems under consideration using a customary framework (see, e.g. \cite{1938London,1992Kagan,Demokritov2006,Bozhko2016,Nazarenko2011}).  
%%%%%%%%%%%%%%%%%%%%%%%%%%%%%%%%%
\subsection{\label{ss:BE-RJ} Bose-Einstein and Rayleigh-Jeans distributions}
 
It is well known from various textbooks (see, e.g. \cite{1980Landau}) that the free evolution of an ideal Bose gas and weakly-interacting wave systems results in the 
BE distribution for particle  (e.g. $^4$He atom) or quasiparticle numbers:
\begin{subequations}\label{distr1}
\begin{equation}\label{1}
    n\Sp{BE}_k=\frac 1{ \exp [( E_k-\mu)/ T]-1} \,,  
\end{equation}
in which $E_k$ is the particle (or quasiparticle) energy, $T$ is the temperature, and $\mu$ is the chemical potential. For non-relativistic particles and quasiparticles with parabolic dispersion law Eq.(\ref{Ek1}), to ensure $T>0$ and $n_k \ge 0$ the value of $\mu$ must be smaller than the minimum of $E_k$: $\mu\le E_0$.

In the low-energy limit, when ($E_k-\mu)<T$, the BE distribution approaches its classical limit, known as the  Rayleigh-Jeans (RJ) distribution\,\cite{1980Landau}:
\begin{equation}\label{RJ}
n_k\Sp{RJ}(T,\mu)=\frac{T}{E_k-\mu}\ .
\end{equation}
In the opposite limit, when $E_k-\mu>T$, the BE distribution\,\eqref{1} becomes exponentially small: 
\begin{equation}\label{4b}
    n\Sp{BE}_k\to \exp [-(  E_k-\mu)/ T] \ .  
\end{equation}
 We can understand the crossover wave number $k_\times$, defined by the equation 
 \begin{equation}\label{4c}
 T=\frac{(\hbar k_\times)^2}{2m}+E_0-\mu\,,
 \end{equation}\end{subequations}
 as a  quantum cutoff of the classical RJ-distribution\,\eqref{RJ}: for $k>k_\times$ it becomes exponentially small according to \Eq{4b}. 
 
Note that the crossover wave number $k_\times$  may exceed the maximal wave number $k\sb{max}\simeq \pi/a_0$, determined by the inter-atomic scale $a_0$ or by some details of the system's dynamics (e.g. by the crossover between the flux- and the thermodynamic equilibrium regimes) as will be clarified below. For simplicity, in the present paper, we assume $k\sb{max}$ to be greater than  $k_\times$.
   %%%%%%%%%%%%%%%%%%%%%%%%%%%%%%%%%
\subsection{\label{ss:Quantum} Quantum nature of a BEC of Bose-atoms and waves}
To stress the quantum-mechanical nature of the BEC in both the ideal Bose gas and the gas of quasiparticles, we shortly recall here some results of the celebrated 1925 paper by Albert Einstein~\cite{Einstein1925in2005}. 
 
Consider the thermodynamic equilibrium in the systems   characterized  by  the total  number $N\sb{tot}$ of $^4$He atoms or quasiparticles and the total energy $E \sb{tot}$, measured for quasiparticles from their energy gap $E_0$: 
\begin{subequations}\label{2B}
  \begin{align}  \label{2Ba}
    N\sb{tot}&= \int \frac{n_{\B k}\Sp{BE} d^d k}{(2\pi)^d} \Rightarrow \frac 1{2\pi^2}\int\limits _0^{k_\times} n_{\B k}\Sp{RJ}k^2 dk \,, \\ \label{2Bb}
    E\sb{tot} &= \int \frac{(\hbar k)^2}{2\, m} \frac{n_{ \B k}\Sp{BE} \, d^d k}{(2\pi)^d} \Rightarrow \frac{\hbar^2}{4\pi ^2 m}\int\limits _0^{k_\times} n_{\B k}\Sp{RJ} \,k^4 d k\ . 
  \end{align} 
\end{subequations}
Here, $n_{\B k}\Sp{BE}$ is the BE distribution, given by \Eq{1}, $d k^d= 4\pi k^2 dk$ in the isotropic 3D case considered here with the dimensionality $d=3$, and $m$ is either the actual mass of $^4$He atoms or the effective mass of quasiparticles. To estimate the integrals in \Eqs{2B}, we replaced $n_{ \B k}\Sp{BE}$ in the rightmost integrals by $n_{ \B k}\Sp{RJ}$ and accounted for the exponential decay of $n_{ \B k}\Sp{BE} $ above the quantum cutoff (i.e. for $k>k_\times$) by introducing the upper limit of integration $k_\times$. 
 
The two relations\,\eqref{2B} allow us to find $T$ and $\mu$ in the final equilibrium state.
However, the direct substitution of $n_k\Sp{RJ}$ into \Eqs{2B} leads to an  immediate problem, known as the ultraviolet catastrophe: both integrals for $N\sb{tot}$ and $E\sb{tot}$ diverge for $k_\times\to \infty$:
\begin{subequations}\label{6}
  \begin{align}  \label{6A}
    N\sb{tot}&\approx \frac T{2\pi^2}\int\limits_0^{k_\times} \frac{k^2 dk}{(\hbar k)^2/(2m)+E_0-\mu}\,, \\ \label{6B}
    E\sb{tot}&\approx \frac{\hbar^2\, T}{4\pi ^2 m}\int\limits_0^{k_\times} \frac{k^4  dk}{(\hbar k)^2/(2m)+E_0-\mu} \ . 
  \end{align} 
\end{subequations}
The only solution is to account for the finite value of the quantum cutoff $k_\times$, i.e. for the quantum character of the problem. 
 
A simple analysis of \Eqs{6} shows that lowering $E\sb{tot}$ with fixed $N\sb{tot}$ leads to smaller $T$, while $\mu $ increases and approaches $E_0$, which is zero for $^4$He atoms or $\hbar \omega_0$ for quasiparticles.  
As $\mu$ reaches $E_0$, \Eq{6A} gives an estimate of the maximal possible number of  excited ``gaseous''  $^4$He atoms or quasiparticles with $E_k>E_0$, which we denote as $N \sb{gas}$:
\begin{subequations}\label{7}
  \begin{equation}\label{7A}
    N \sb{gas} \approx \frac{k_\times^3}{4\pi^2}\,,\quad k_\times = \frac{\sqrt {2 m T}}\hbar\ .
  \end{equation}
The corresponding parameter $E\sb{tot}$ we denote as $E\sb{gas} $. According to \Eq{6B} we find:
  \begin{equation}\label{7B}
    E\sb{gas}\approx   N\sb{gas}\frac{(\hbar k_\times)^2}{6 m} \approx \frac{ \pi^{4/3} \hbar ^2}{3 \cdot 2^{1/3}m} \big ( N\sb{gas}\big )^{5/3} \ .
  \end{equation}
\end{subequations}

For $E\sb{tot}<E\sb{gas}$,  the number of ``gaseous'' $^4$He atoms or quasiparticles at the excited energy levels $N\sb{gas}$ becomes smaller than their total number $N \sb{tot}$. What happens with their excess number $N\sb{tot}-N\sb{gas}$? The answer was given by Einstein in \REF{Einstein1925in2005}: the excess  $^4$He atoms (and quasiparticles, as we understand now) occupy only ONE level with the minimal energy, independently of the size of the system, forming a BEC, in which the number of BE condensed atoms or quasiparticles  $N\Sb{BEC}=N\sb{tot}-N\sb {gas} $ is macroscopically large. All these BE condensed quasiparticles belong to the basic quantum state with its wave function coherent over the entire size of the system, owing to the fundamental principle of quantum mechanics of the indistinguishability of identical quasiparticles: the particles (or quasiparticles) with zero and natural spins can occupy any quantum state without limitation of their occupation number.  
 
The existence of the quantum cutoff and the indistinguishability of identical (quasi)particles are necessary
conditions for the BE condensation of  Bose atoms and quasiparticles.
Therefore, the phenomenon of the BE condensation of $^4$He atoms and other Bose atoms, as well as magnons and other quasiparticles, has a fundamentally quantum nature. 

%%%%%%%%%%%%%%%%%%%%%%%%%%%%%%%%%
\subsection {\label{Q2D-eq} Quasi-BEC in two-dimensional systems}
In the 2D isotropic case, $d^2k= 2\pi k dk $ and integrals for $N\sb{tot}$ and $E\sb{tot}$, similar to \Eqs{6}, take the form:
 \begin{subequations}\label{2D}
  \begin{align}\label{2DA}\begin{split}
     N\sb{tot} \approx  & \frac T{2\pi }\int\limits _0^{k_\times}  \frac{k   dk}{(\hbar k)^2/(2m) -\delta \mu}  
     =   \frac{k_\times^2}{4\pi}  \ln  \! \Big (1 - \frac T {\delta \mu}\Big)   \\ \approx & \frac{k_\times^2}{4\pi} \ln  \!\Big ( \frac T {|\delta \mu|}\Big) =   \frac{k_\times^2}{2\pi} \ln  \!\Big ( \frac  {k_\times}{k\sb{min} }\Big)\,,  
  \end{split}   \\  
     \begin{split} 
     E\sb{tot}\approx & \frac{\hbar^2\, T}{8\pi ^2 m}\int\limits _0^{k_\times}  \frac{k^ 3  dk}{(\hbar k)^2/(2m) -\delta \mu} \\
      =&  \frac{  k_\times^2}{4\pi}\Big [T +\delta \mu \ln  \Big (1 + \frac T {\delta \mu}\Big)\Big ]\approx  \frac{T  k_\times^2}{4\pi} \,, 
  \ \mbox{where}   \end{split}   \\  
     \begin{split} \delta \mu\=&  \mu- E_0<0\,, \ \mbox{and} \  k\sb{min}\=\frac{\sqrt{2 m |\delta \mu|}}{\hbar} \ .
    \end{split}   \label{2DB} 
     \end{align}  
 \end{subequations} 
As $\delta \mu\to 0$, $N\sb{tot}$ becomes logarithmically large, i.e., any large number of quasiparticles can occupy excited levels with $k>0$. Therefore, BE condensation never happens in unbounded 2D media.   

According to \Eqs{Ek} and \eqref{RJ}, for $  \mu =E_0= \hbar \omega _0$ the wave distribution diverges at $k=0$:  
\begin{subequations}\label{9}
 \begin{equation}
    n_k\Sp{RJ}=\frac{T}{\hbar \omega_0 (ak)^2}=\frac{2 m T}{(\hbar k)^2} 
   \end{equation}
   and formally $N\sb{tot}=\infty$. Nevertheless, when 
    $N\sb{tot}\to \infty$ but still finite, $  \mu  \to \hbar \omega_0$, and $k\sb{min}\to 0$, the coherence length of the quasiparticles
 \begin{equation}\label{9B}
  \ell\simeq \pi/ k\sb{min}  
 \end{equation}
increases and finally reaches the sample size. In other words, the wave system becomes coherent across the entire sample and can be practically considered as a BEC. Nevertheless, to be formally rigorous, we will refer to this system as quasi-BEC\,\cite{1992Kagan}.
 \end{subequations} 
 
%%%%%%%%%%%%%%%%%%%%%%%%%%%%%%%%%  
\section{\label{s:FE-WP}Flux equilibrium with weak pumping: quasi-equilibrium regime} 
In this Section we consider an isotropic system with a parabolic dispersion law \Eq{Ek} and relatively weak pumping, such that in the stationary case the system can be considered close to the thermodynamic equilibrium.  Then, similar to the equilibrium regime, main contributions to all $N\sb{tot}$- and $E\sb{tot}$-integrals, given by \Eqs{2B}, come from the range $k<k_\times$, i.e., below the crossover between the quantum and classical scales. To simplify the appearance of our results, we approximate, analogous to the previous \Sec{ss:Quantum}, the BE distribution~\eqref{1} in this range by the Rayleigh-Jeans distribution~\eqref{RJ}. 

The idea of the theoretical analysis of the BE condensation in the quasi-equilibrium regime is simple. By balancing the rate of the quasiparticle input with the rate of their loss, we will find their total number $N\sb{tot}$ in the flux-equilibrium regime. A similar analysis of the rate equation for the energy allows us to find the total energy of the system $E\sb{tot}$. Because the system is assumed to be close to the thermodynamic equilibrium in the vicinity of the energy minimum, we can find an effective temperature $T\sb{eff}$, describing a local Rayleigh-Jeans distribution in this region. In turn, this allows us to find the number of excited quasiparticles $N\sb{gas}$, occupying energy levels $E_k>E_0$. If  $N\sb{gas}$ turns out to be smaller than $N\sb{tot}$, the excess quasiparticles create the BEC: $N\sb{tot}-N\sb{gas}=N\Sb{BEC}$. Otherwise, there is no BEC. 

%%%%%%%%%%%%%%%%%%%%%%%%%%%%%%%%%
\subsection{\label{ss:3D-BVEC} BEC in 3D-systems}
To study BE condensation in 3D systems along the lines of the suggested procedure, consider now the continuity equation for the quasiparticle numbers $n_{\B k}$ with a damping frequency $\gamma$ (taken  for simplicity as $\B k$-independent) and a source (influx) of quasiparticles $f$ at a surface of the sphere of radius $k\sb f$:

\begin{equation}\label{Bal}
  \frac{ k^2 }{2\pi^2} \frac{\partial n_k}{\partial t}+ \frac{\partial \eta _k}{\partial k} =-  \frac{ \gamma k^2 }{2\pi^2}  \, n_k+ \frac{ k^2\sb  f f}{2\pi^2} \delta (k-k \sb f)\ .
\end{equation}
Here $\eta_k$ is the flux of quasiparticles in the 1D-space of $k=|{\B k}|$ (the angle-averaged 3D $\B k$-space),

The rate equation for the energy (measured from $E_0$)is obtained by multiplying \Eq{Bal} by $\hbar \delta \omega_k = \hbar\, \omega_0 a^2 k^2$.  Integrating the result, we obtain for the total energy in 3D:
\begin{subequations}\label{3D-Bal}\begin{equation}\label{3D-E}
E\sb{tot}\Sp{3D}=\frac{\hbar \omega_0 a^2 f k\sb f^4}{2\pi ^2 \gamma}\ .
\end{equation}
Now, using the definition \Eq{mass} and in analogy to \Eq{6B} with the effective temperature $T=T\sb{eff}$ and the effective mass $m=m\sb{eff}$, as well as assuming the presence of the BEC (i.e. $\mu=E_0$), we rewrite $E\sb{tot}$ as:
  \begin{align}
    \begin{split}\label{8B}
     E\sb{tot} \Sp{3D} =&  \frac{  T\sb{eff}    k_\times^3}{6\pi^2}\ . 
    \end{split}
  \end{align}
\end{subequations} 
Equations\,\eqref{3D-Bal} allows us to find $T\sb{eff}$. Using it to find, with the help of \Eq{6A}, the total number of exited (gaseous) quasiparticles with $k>0$ we obtain:
   \begin{equation}\label{gas}
   N\sb{gas}\Sp{3D}= \frac{  T\sb{eff} \, k_\times }{2 \pi^2 \hbar \, a^2 \, \omega_0}= \frac{3 f k\sb f^4}{2\pi^2k_\times^2 \gamma}\ .
   \end{equation}
On the other hand, the total number of quasiparticles $N\sb{tot}$ can be found by integration of the particle rate \Eq{Bal}:
   \begin{equation}\label{Ntot}
 N\sb{tot}\Sp{3D}=\frac{f\, k\sb f^2}{2\pi^2 \gamma}\ .
   \end{equation}
  
The excess of $N\sb{tot}\Sp{3D}$ over $N\sb{gas}$ is the number of BE condensed quasiparticles: 
  \begin{align}\label{13}  
      N\Sb{BEC}\Sp{3D}& = N\sb{tot}\Sp{3D}-N\sb{gas}\Sp{3D}=\frac{f k\sb{f}^2}{2\pi^2\,\gamma}\Big( 1- \frac{ k\sb f^2}{k\sb{cr}^2}\Big)\,, \quad k\sb{cr}^2\= \frac{k_{\times}^2} 3  \ .
  \end{align}
Recall that by setting $\mu=E_0$ we assumed that the BEC is formed, i.e., $N\Sb{BEC}\Sp{3D}>0$. 
We see that the BEC appears only if the spectral location of the quasiparticle influx $f$ is below the critical value $k\sb{cr}$, which is independent of the value of $f$. However, if the condition $k\sb f<k\sb{cr} $ is fulfilled, the particle number in the BEC is proportional to $f/\gamma$. The numerical factor 3 in \Eq{13} is a consequence of the simplifying assumption that $\gamma_k$ is $k$-independent.   
  
The requirement of the smallness of $k\sb{f}$ for BE condensation has a simple physical meaning. In our model with monochromatic pumping [at a single frequency $\omega(k\sb f)$] of the energy (counted from $E_0$) and the quasiparticles, their influxes, $P_{_E}$ and $P_{_N}$ are related: $P_{_E}= \hbar[\omega(k\sb f)-\omega(0)] P_{_N}$. This means that at the constant quasiparticle influx $P_{_N}$ (and, consequently, constant $N\sb{tot}\Sp{3D}$), the energy influx $P_{_E}$, being proportional to $k\sb f^2 P_{_N}$, decreases with decreasing $k\sb f$. Accordingly, $E\sb{tot}$ and $T\sb{eff}$ also decrease with decreasing  $k\sb f$. Clearly, the wave system with the constant $N\sb{tot}$ will unavoidably experience BE condensation when $k\sb f$ (and consequently $T\sb{eff}$) become smaller and smaller. On the other hand, for large $k\sb f$ in the hot system there will be no BEC. Hence, there is a critical value $k\sb{f}=k\sb{cr}$, \Eq{13}, at which BE condensation happens. 
  
\subsection{\label{ss:2D-BVEC} Quasi-BEC in 2D-systems} 
As we discussed in \Sec{Q2D-eq}, one expects the appearance of a quasi-BEC in a finite-size 2D-space, say in a square domain $L\times L$. Assuming for concreteness periodical boundary conditions, we end up with a discrete $\B k$-space, in which $k_x=\pm 2\pi n_x/L$ and  $k_y=\pm 2\pi n_y/L$ with $n_x$ and $n_y=0\,, 1\,, 2\,, \dots L$\ . The wave vector $\B k=0$, i.e. $k_x=k_y=0$, can be considered as the position of the quasi-BEC, while the rest of the $\B k$-space can be roughly approximated as a continuous $\B k$-space, restricted by the inequality $k>\widetilde k\sb{min}\approx 2\pi/L$. If so, the quasiparticle distribution  in 2D case reads:
\begin{equation}\label{14}
    n _k=\frac {T \theta (k\sb{min})}  { \hbar \, \omega_0 (a\,k)^2  } + (2\pi)^2N\Sb{BEC}\delta^2{(\B k)}\,, \quad k<k_\times\ ,  
\end{equation}  
where $\theta (k\sb{min})$ is the Heaviside step function. 

The 2D version of the quasiparticle rate equation \eqref{Bal} is:
\begin{equation}\label{Bal1}
  \frac{ k }{2\pi} \frac{\partial n_k}{\partial t}+ \frac{\partial \eta _k}{\partial k} =-  \frac{ \gamma k }{2\pi}  \, n_k+ \frac{ k\sb f f}{2\pi} \delta (k-k\sb f)\ .
\end{equation}
Multiplying the stationary version of  this equation   by $\hbar\, \omega_0 a^2 k^2$ and integrating  over $d k$ gives  for the total energy  
\begin{subequations}\label{2D-Teff}
 	\begin{align} \label{17A}
 E\sb{tot}\Sp{2D}= \frac {\hbar \omega_0 f \, a^2 k\sb f^3}{2\pi \gamma }  \,,
    \end{align}
similar to \Eq{3D-E}. By analogy with  \Eq{6B},  $E\sb{tot}\Sp{2D}$ in terms of the effective temperature reads    
	\begin{equation}\label{Etot-2D}
E\sb{tot}\Sp{2D}= \frac{T\sb{eff}k_\times ^2}{4\pi}\ .
	\end{equation}
\end{subequations} 
Similar to the 3D case, these equations allow one to find $T\sb{eff}$ and, by the help of a 2D version of \Eq{6A}, the total number of gaseous quasiparticles in 2D:
\begin{align} 
	\begin{split} \label{18}
   N\sb{gas}\Sp{2D} &=\frac{ T\sb{eff} }{2 \pi \hbar \, \omega_0 a^2}\ln \Big(\frac {k_\times}{\widetilde k\sb{min}} \Big )  \\
&=\frac{ f k\sb f^3 }{ k_\times^2 \gamma  }\ln \Big(\frac {k_\times}{\widetilde k\sb{min}} \Big )\,,\qquad   \widetilde  k\sb{min}\approx \frac{2\pi}L \,,
	\end{split}  
\end{align} 
instead of \Eq{gas} for the 3D case.  

Now, integrating the stationary \Eq{Bal1} over $k$, one gets a new  equation for the total particle number $N\sb{tot}\Sp{2D}$ which gives,
   similar to \Eq{Ntot} for $N\sb{tot}\Sp{3D}$:
\begin{equation}\label{Ntot-2D}
  N\sb{tot}\Sp{2D}=\frac{f k\sb f}{2\pi \gamma}\ .
\end{equation}

As in the 3D case, the excess of $N\sb{tot}\Sp{2D}$ over $N\sb{gas}\Sp{2D}$ gives the number of BE condensed quasiparticles: 
\begin{subequations}
  \begin{align}\label{16A} 
    N\Sb{BEC}\Sp{2D} &=  N\sb{tot}\Sp{2D}-N\sb{gas}\Sp{2D}  = \frac{f k\sb{f} }{2\pi \,\gamma}
    \Big( 1- \frac{ k_f^2}{k\sb{cr}^2}\Big) \,,  \\ \label{16B}
    k\sb{cr}^2 &\= k_\times^2\Big / \Big [ 2 \,\ln \Big (\frac{k_\times}{\widetilde k\sb{min}}\Big ) \Big ] \ .
  \end{align}
We see that also in 2D the BEC appears only if the position of the quasiparticle influx $f$ is below the critical value $k\sb{cr}$, now given by \Eq{16B}. As before, $k\sb{cr}$ is independent of the value of this influx $f$ and $k\sb{cr}< k_\times$. Also, similar to the 3D case, when $k_f<k\sb{cr}$, the particle number in the BEC is proportional to $f/\gamma$. 
\end{subequations} 
%%%%%%%%%%%%%%%%%%%%%%%%%%%%%%%%%

\section{\label{s:FE-SP}Flux equilibrium with strong  pumping: scale-invariant regimes} 
\subsection{\label{ss:KE}
Kinetic equation and damping frequency}

A consistent description of the evolution of  an overpopulated wave system towards the formation of a BEC may be achieved in the framework of the theory of weak wave turbulence~\cite{Zakharov1992,Nazarenko2011}. The main tool of this theory is a kinetic equation (KE) for the occupation numbers $n(\B k)$ of quasiparticles:  
\begin{equation}\label{KE}
\frac{\partial n_{\B k}}{\partial t}=   \mathrm{St} (\textbf{\textit{k}}, t)\ .
\end{equation}
The collision integral $\mathrm{St} (\textbf{\textit{k}}, t)$ may be found by various ways~\cite{Zakharov1992,Nazarenko2011}, including the Golden Rule, widely used in quantum mechanics~\cite{Landau1977}. In the case  of the three-wave    decay
\begin{subequations} \label{3w}
	\begin{equation}\label{dec}
 \omega_{\B k} =\omega_{\B 1} +\omega_{\B 2}\,, \quad
 \B k =\textbf{\textit{k}}_1+\textbf{\textit{k}}_2\,, 
	\end{equation}
and confluence  processes
	\begin{equation}\label{dec2}
 \omega_{\B k} +\omega_{\B 1} =\omega_{\B 2}\,, \quad
 \B k +\textbf{\textit{k}}_1=\textbf{\textit{k}}_2\,, 
	\end{equation}
\end{subequations} 
the collision integral takes the form\,\cite{Zakharov1992,Nazarenko2011}:
\begin{subequations}\label{3-St}  
	\begin{align}
		\begin{split}
 	^3\mbox{St} (\textbf{\textit{k}} ,t)  =  \pi  &\int    d \B k_1 d\B k_2   \Big [\frac 12 |V_{\B k}^{ \B 1 \B 2}|^2 \, \mu _{\B k}^{ \B 1 \B 2}\\  \times  &\delta( \B k -\B k_1 -\B k_2)   
  	 \delta (\omega_{\B k}  -\omega_{\B 1} -\omega_{\B 2} )   \, 
 		\end{split}   	 \\  
 + |V_{\B 2}^{\B k \B 1}|^2\, \mu _{\B 2}^{\B k \B 1} \,   &\delta( \B k_2 -\B k_1-\B k)   	 \delta (\omega_{\B k_2}  -\omega_{\B 1} -\omega_{\B k} )  \ .
	\end{align} 
\end{subequations} 
Here $\omega_j\= \omega_{k_j}\=\omega(k_j)$, $n_j \= n_{{\bf k}_j}$, $V_{\B k}^{\B 1 \B 2}\= V(\B k , \B k_1 \B k_2)$   is the  3-wave interaction amplitude and  $\mu _{\B k}^{\B 1 \B 2} \= n_{\B 1}n_{\B 2}-   n_{\B k}(  n_{\B 1}+ n_{\B 2})$. 
 
If the three-wave processes\,\eq{3w} are suppressed or forbidden, the main role is played by four-wave
 $2\Leftrightarrow  2$ 
%$2\leftRight \Leftrightarrow  2$
processes 
\begin{subequations}\label{2-2-proc}
	\begin{equation}\label{2-2}
\omega_\textbf{\textit{k}}  +\omega_{\B 1} =\omega_{\B 2} +\omega_{\B 3}\,, \quad
\textbf{\textit{k}}+\textbf{\textit{k}}_1=\textbf{\textit{k}}_2+\textbf{\textit{k}}_3 \ .
	\end{equation}
In this case, the  collision integral reads \cite{Zakharov1992,Nazarenko2011}:
	\begin{align}
 \label{4-St}  
 		\begin{split}
 	^4\mbox{St} (\B {k}  ,t)=  & \frac{\pi}{4}
   \int  d\B {k}_1 d\textbf{\textit{k}}_2 d\textbf{\textit{k}}_3 \,    \delta(\textbf{\textit{k}}+\textbf{\textit{k}}_1-\textbf{\textit{k}}_2-\textbf{\textit{k}}_3)\\ 
 &\times 	 \delta (\omega_\textbf{\textit{k}}  +\omega_{\B 1} -\omega_{\B 2} -\omega_{\B 3} ) \,   
 	|W _{\textbf{\textit{k}} \B 1}^{\textbf{2} \textbf{3}}|^2  \, \\  
 &	\times  [n_\textbf{2}   n_\textbf{3}  (n_\textbf{\textit{k}}    + n_\textbf{1}  )- n_\textbf{\textit{k}}   n_\textbf{1}  ( n_\textbf{2}  + n_\textbf{3} ) ]\ .
		\end{split}
	\end{align}  
\end{subequations}
Here  $W _{\textbf{\textit{k}} \B 1}^{\textbf{2} \textbf{3}}=W_{\B k, \B k_1}^{\B k_2,\B k_3}  =W(\B k, \B k_1; \B k_2,\B k_3)$ is the  four-wave interaction amplitude.  

The kinetic equations\,\eqref{KE}, \eqref{3-St} and \eqref{4-St} have a stationary thermodynamic equilibrium solution in the form of the RJ distribution\,\eqref{RJ}. To describe the evolution of the system close to the RJ distribution, \Eq{KE} can be approximately reformulated as follows,
\begin{equation}\label{ev}
\frac{\partial n_{\B k}}{\partial t}= \gamma_{\B k}\big[n\Sp{RJ}_{\B k} -  n_{\B k} \big ]\
\end{equation}
where $\gamma_{\B k}$ is proportional to the part of the collision integral that explicitly includes $n_{\B k}$.
In the case of four-wave processes\,\eqref{2-2} with the collision integral\,\eqref{4-St} we find:
\begin{subequations}\label{balance}
	\begin{align}
		\begin{split}\label{gamma} 
 \white{.}\hskip -.4  cm	^4\gamma_{\B k}& =  \frac{\pi}{4} \!\! \int \!\! d\textbf{\textit{k}}_1 d\textbf{\textit{k}}_2 d\textbf{\textit{k}}_3  	|W _{\textbf{\textit{k}} \B 1}^{ \textbf{2} \textbf{3}}|^2 \delta(\textbf{\textit{k}}+\textbf{\textit{k}}_1-\textbf{\textit{k}}_2-\textbf{\textit{k}}_3)\\ 
 &\times 	 \delta (\omega_\textbf{\textit{k}}  +\omega_1 -\omega_2 -\omega_3 )  
   [ n_{\B 1}  ( n_\textbf{2}  + n_\textbf{3}) - n_{\B 2} n_{\B 3} ]\ .
		\end{split} 
	\end{align} 
 According to \Eq{ev}, near the equilibrium 
	\begin{equation}\nonumber \label{rel}
    n_{\B k}-n\Sp{RJ}_{\B k}  \propto \exp [ - \ ^4\gamma_{\B k}t] \ .
	\end{equation}
Therefore, $^4\gamma_{\B k}$ \eqref{gamma} has a meaning of the damping (or relaxation) frequency in the four-wave scattering processes \eqref{2-2}. Close to and at equilibrium it is positive $^4\gamma_{\B k}>0$.

In general, the kinetic \Eq{KE} with the collision term\,\eqref{4-St} can be written as follows:
	\begin{equation}\label{ev-2}
\frac{\partial n(\B k)}{\partial t}=  \Phi(\B k)-\gamma(\B k)  n(\B k)  \,,
	\end{equation}
where $n(\B k)\=n_{\B k}$,  $\gamma(\B k)=\, ^{4 \!}\gamma(\B k)$ is given by \Eq{gamma} and 
the source term has the form
	\begin{align}
 \label{incom}  
		\begin{split}
 	 \Phi(\B k)  =&  \frac{\pi}{4}
   \int  d\B {k}_1 d\textbf{\textit{k}}_2 d\textbf{\textit{k}}_3 \,   \delta(\textbf{\textit{k}}+\textbf{\textit{k}}_1-\textbf{\textit{k}}_2-\textbf{\textit{k}}_3)\\ 
 &\times 	 |W _{\textbf{\textit{k}} \B 1}^{\textbf{2} \textbf{3}}|^2 \delta (\omega_\textbf{\textit{k}}  +\omega_{\B 1} -\omega_{\B 2} -\omega_{\B 3} ) \,   
 	    n_\textbf{1}   n_\textbf{2}   n_\textbf{3}    \ .
		\end{split}
	\end{align} 
\end{subequations} 

In thermodynamic equilibrium  $n_{\B j}=n_{\B j}\Sp{RJ}$, $\Phi(\B k)=\gamma(\B k)n_{\B k}\Sp{RJ}$ and \Eq{ev-2} coincides with \Eq{ev}, as expected.

\subsection{\label{ss:Scale}
Scale-invariant solutions of the kinetic equation}
Weak wave turbulence theory\,\cite{Zakharov1992,Nazarenko2011} also allows us to find  stationary flux  solutions of the KE in the isotropic scale-invariant case, for which the wave frequency depends only on $k =|\B k|$  and the interaction amplitude $W _{\textbf{\textit{k}} \B 1}^{\textbf{2} \textbf{3}}$ is a homogeneous function,
\begin{align}\label{scaling}
	\begin{split}
\omega_{\B k} & =\omega_k\propto k^\alpha\,, \qquad  
W_{\sigma \B k_1,\sigma  \B k_2}^{\sigma  \B k_3,\sigma  \B k_4 }   = \sigma ^m W_{   \B k_1,  \B k_2}^{   \B k_3,  \B k_4 }.
%\simeq W_0 (k_1,k_2,k_3,k_4)^{m/4}\ .
	\end{split}
\end{align} 
Here, $\alpha$ is the frequency scaling index (for example, for magnons at the beginning of the exchange-dominated dispersion branch $\alpha=2$), $m$ is the four-wave interaction-amplitude scaling index and $\sigma$ is a positive constant. For simplicity, let us take
\begin{equation}
 \label{W}
 W _{\textbf{\textbf{1}} \B 2}^{\textbf{3} \textbf{4}} = W_0 a^m (k_1 k_2 k_3 k_4)^{m/4}\ , 
\end{equation}
where $W_0$ is a constant. Up to now, our analysis has a general character, applicable to any nonlinear wave system. Below, having in mind the comparison of our predictions with spin waves in a ferromagnetic material, we choose the interaction parameters typical for this system for further discussion. In the ferromagnetic material, at sufficiently large $k$, the exchange interaction with $m=2$ is dominant. In the low-$k$ range, the dipole-dipole interaction dominates, with $m=0$.

The scaling solutions (up to a dimensionless prefactor) read 
\begin{subequations}\label{sol28}
	\begin{eqnarray}\label{sol28A}
n_\varepsilon(k)&\simeq &\frac {\varepsilon ^{1/3}}{W_0 ^{2/3}(a\,k)^{x_\varepsilon}}\,, \quad x_\varepsilon= d+ \frac{2m}3\,, \\
\label{sol28B}
 n_\eta (k)  &\simeq& \frac {\eta  ^{1/3}}{W_0 ^{2/3}(a\,k)^{x_
\eta }}\,, \quad x_\eta = d+ \frac{2m-\alpha }3\ .
	\end{eqnarray}
\end{subequations}
Here $\varepsilon$ and $\eta $ are the energy and the quasiparticle (magnon) number fluxes, and $d$ is the dimensionality of the space.

  %%%%%%%%%%%%%%%%%%%%%%%%%%%%%%%%%
 
\subsection{Directions of the fluxes and realizability  of the  flux solutions in ferromagnets}

Following the Fj\o{}rtoft argument~\cite{Fjortoft1953}, one can show  (see, e.g.  \cite{Zakharov1992,Nazarenko2011}) that the energy flux solution \eqref{sol28A} is oriented toward large $k$ (``direct energy cascade''), while the quasiparticle-flux  solution \eqref{sol28B} flows toward small $k$ (``inverse particle cascade''). This conclusion is based on the analysis of the energy and quasiparticle number balance in the stationary, scale-invariant, isotropic situation, in which the energy and the quasiparticles are pumped around some intermediate wave number $k\sb f$ and dissipate at both very small $k\Sb <$- and very large $k\Sb>$-numbers:  $k\Sb<\ll k\sb f$, $  k\Sb>\gg k\sb f $. 

Now we are going to verify that the scale-invariant solutions~\eqref{sol28}  do have the directions of the energy and magnon number fluxes that agree with the Fj\o{}rtoft argument. 

\begin{figure}
    \includegraphics[width=1\columnwidth]{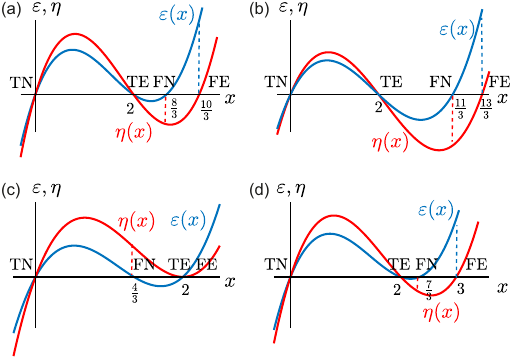}
    \caption{Schematic presentation of the particle flux $\eta (x)$ (in red) and the energy flux $\varepsilon(x)$ (in blue) as a function of spectral index $x$. (a) exchange-dominated interaction, 2D; (b) exchange-dominated interaction, 3D; (c) dipole-dipole dominated interaction, 2D; (d) dipole-dipole dominated interaction, 3D.
            }  
    \label{f1}
\end{figure}

For that, we analyze the behavior of these fluxes for the distributions $n(k)\propto k^{-x}$ with an arbitrary value of $x$. We expect that for a very steep spectrum, the fluxes will act to change it toward the equilibrium spectra. Therefore, for large and positive $x$, given that $n(k)$ decreases sharply toward larger wave numbers, we expect both $\varepsilon$ and $\eta$ fluxes to be positive, i.e., directed toward large $k$. On the other hand, for large negative $x$, when the spectra are growing toward high wave numbers, we expect $\varepsilon, \eta <0$. Furthermore, both fluxes will be zero for both thermal equilibrium exponents $x\Sp{TE}_\eta =2$ (because $n\Sb{RJ}\propto 1/\omega_k$) and $x\Sp{TE}_\varepsilon =0$.
In addition, the  flux of magnons $\eta (x)$ vanishes for the pure energy flux spectrum with exponent  $x_\varepsilon$ given by \Eq{sol28A}, and the energy flux $\varepsilon(x)$ vanishes for
the pure particle flux exponent $x_\eta $, given by \Eq{sol28B}.
By continuity, the signs of both fluxes for all $x$ are fully determined by their signs at infinity and the locations of their zero crossings. The fluxes vary in the manner schematically shown in Figs.\,\ref{f1}(a) 
and (b) for the 2D and 3D case, respectively, with the exchange-dominated interaction with $m=2$, and Figs.\,\ref{f1}(c) and (d) for the 2D and 3D case, respectively, with the dipole-dipole dominated interaction with $m=0$ \cite{Lvov1993}.

First, we consider the exchange-dominated case in ferromagnets ($m=2$) for which
\begin{subequations}
	\begin{eqnarray}
 x_\varepsilon=\frac{10}3\,, && \quad x_\eta =\frac{8}3\,,
 \quad  \mbox{for} \ d=2\,,\\
 x_\varepsilon=\frac{13}3\,, && \quad x_\eta =\frac{11}3\,,
 \quad  \mbox{for} \ d=3\ .
	\end{eqnarray}
\end{subequations}
We see in Figs.\,\ref{f1}(a,b)  that in both 2D and 3D at the spectral index $x=x_{\varepsilon}$, corresponding to the pure energy flux ($\frac{10}3$ and $\frac{13}3$ respectively), the energy flux is positive, $\varepsilon>0$. Similarly, we see that for $x=x_{\eta }$, corresponding to the pure flux of magnons ($\frac{8}3$ and $\frac{11}3$ respectively), the   flux of magnons is negative, $\eta <0$. These findings are in full agreement with the Fj\o{}rtoft prediction.

In the dipole-dipole interaction-dominated case, when $m=0$, 
\begin{subequations}
	\begin{eqnarray}\label{25A}
 x_\varepsilon=2\,, && \quad x_\eta =\frac{4}3\,,
 \quad  \mbox{for} \ d=2\,,\\ \label{25B}
 x_\varepsilon=3\,, && \quad x_\eta =\frac{7}3\,,
 \quad  \mbox{for} \ d=3\ .
	\end{eqnarray}
\end{subequations}
These exponents are the same as in the Nonlinear Schr\"{o}dinger (NLS) equation, studied in Refs.\,\cite{Dyachenko1992,Nazarenko2011,Skipp2020}. Consequently, our schematic representation of the fluxes \Fig{f1}(c,d) looks similar to Fig.\,1(c,d) in \REF{Skipp2020}. For completeness, the analysis that led to the conclusions made in \REF{Skipp2020}, is briefly reproduced below. 
 
In 3D, $\varepsilon$ is positive at $x_\eta $ and $\eta $ is negative at $x_\varepsilon$ in agreement with the Fj\o{}rtoft argument. We therefore can expect that in 3D the Kolmogorov-Zakharov flux cascades with exponents~\eqref{25B} are possible. A more detailed analysis~\cite{Dyachenko1992,Nazarenko2011} shows that the inverse particle cascade Kolmogorov-Zakharov spectrum is indeed realized, while the direct energy cascade is marginally nonlocal and the respective spectrum must be modified by the logarithmic corrections.

As seen in \Fig{f1}(d), in spectra \eqref{sol28} the quasiparticles cascade is directed to large $k$ and the energy cascade is zero. This contradicts the robust Fj\o{}rtoft-type analysis based on the energy and the quasiparticles number balance for the situation when the energy and quasiparticles are pumped around some intermediate wave number $k\sb f$ and dissipate in both ranges of very small $k\Sb <$- and very large $k\Sb>$-numbers.
The contradiction may be resolved if, instead of the pure scaling spectra \eqref{sol28}, the inverse quasiparticles  and the direct energy cascades are realized by spectra with a shape close the thermodynamic RJ equilibria~\eqref{RJ} with small corrections which take care of the magnon and energy fluxes toward small and large $k$, respectively.
%%%%%%%%%%%%%%%%%%%%%%%%%%%%%%%%%%%%%%%%

\begin{figure}[b]
    \includegraphics[width=1\columnwidth]{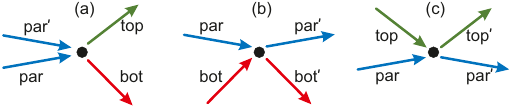}
    \caption{\label{F:KI-diagrams}(a) Interaction processes \Eq{cons1} leading to the kinetic instability. (b) and (c), scattering processes \eqref{cons2}, leading to the widening of the frequency distribution of the bottom and top quasiparticles.
            }   
\end{figure}   

\subsection{Transition from 3D- to 2D-cases in thin films}
In thin films, we chose the direction $z$ orthogonal to the film surface. The corresponding wave vector is $k_{z}=  \pi n_z  /\Delta$, where $\Delta$ is the film thickness. Accordingly, the wave frequency is also quantized, and the frequency of the fundamental mode with $n_z=1$ is separated from the frequency of the next mode with $n_z=2$, etc. This results in the appearance of a crossover wave number $k_{2\leftrightarrow3}$ between the 2D and 3D regimes in the flux solutions. For $k\gg k_{2\leftrightarrow3}$, the 3D flux solutions can be realized, while for $k\ll k_{2\leftrightarrow3}$ the energy and quasiparticles exchange between waves with different $n_z$ are strongly suppressed, or even forbidden. Roughly speaking, for $k_{2\leftrightarrow3}$ the frequency gap between waves with neighboring $n_z$ is of the order of the interaction frequency (or damping frequency) of waves with $k\sim k_{2\leftrightarrow3}$. 
  
For the problem at hand it means the following: if the pumping wave number strongly exceeds the crossover $k\sb f\gg k_{2\leftrightarrow3}$, there exists a direct energy flux toward large $k$ for $k> k\sb f$, whereas in the intermediate range $ k_{2\leftrightarrow3} <k< k\sb f$, an inverse particle flux toward small $k$ is realized. For small wave numbers $k<k_{2\leftrightarrow3}$, the wave system falls into the 2D regime, in which the scale-invariant flux solution cannot be realized. Instead, it approaches a solution close to the thermodynamic equilibrium with small deviations ensuring the required particle flux. 
%%%%%%%%%%%%%%%%%%%%%%%%%%%%%%%%%

\section{\label{s:KI} Ultra-strong pumping: Kinetic instability and BEC}
In this Section, we consider ultra-strong parametric pumping by an external monochromatic field of frequency $\omega\sb{pump}$, exciting a very intense package of quasiparticles near the resonant frequency $\omega\sb {par}=\omega\sb{pump}/2$ according to \Eq{cl}, see \Fig{f:0}. As predicted  in Ref.\,\cite{Lavrinenko1981}, 
the scattering process with the conservation law 
\begin{equation}\label{cons1}
\omega (\bm k\sb{par}) + \omega (\bm k'\sb{par}) =\omega (\bm k\sb{bot})+\omega (\bm k\sb{top})\,,
\end{equation}
shown in \Fig{F:KI-diagrams}(a) decreases the damping frequency of the bottom and top waves $\gamma (\B k\sb {bot})$ and $\gamma (\B k\sb {top})$ such that they may become negative. If so, the cascade processes, discussed in \Sec{ss:Scale}, may be augmented by a direct transfer of the parametrically injected quasiparticles to the lowest energy states, creating a dense cloud of incoherent ``bottom'' quasiparticles formed close to the BEC point \cite{Kreil2018} and a similar cloud at the high energy state (``top quasiparticles'').
   
The linear stage of this phenomenon, referred to as kinetic instability\,\cite{Lavrinenko1981,Lvov1981c,Melkov1991,Melkov1994}, is considered below in section\,\ref{ss:LKI}. The exponential growth of the number of the bottom and top quasiparticles due to kinetic instability alters the damping of the parametric ones. This process is discussed in \ref{ss:S-theory}. In its turn, two other scattering processes, shown in \Figs{F:KI-diagrams}(b) and (c), involve parametric, bottom, and top waves
\begin{align}
	\begin{split}\label{cons2}
\omega (\bm k\sb{par}) +  \omega (\bm k \sb{bot})& = \omega (\bm k'\sb{par})+ \omega (\bm k'\sb{bot})\,,\\
\omega (\bm k\sb{par}) +  \omega (\bm k \sb{top})&= \omega (\bm k'\sb{par})+ \omega (\bm k'\sb{top})\ .
	\end{split}
\end{align}
It widens the frequency distribution of the bottom and top quasiparticles, as described in \ref{ss:scat}. Combining all processes together, we formulate the nonlinear description of the kinetic instability in \ref{ss:NLT}.  

Although the kinetic instability was first discovered in a system of parametrically excited magnons \cite{Lavrinenko1981}, it is a general physical phenomenon in nonlinear wave systems. To stress this generality and to clarify the underlying phenomena, we describe it here for the isotropic homogeneous wave system with the parabolic dispersion law\,\eqref{3}.  
We postpone the discussion of the specific features related to the anisotropic spectrum of magnons, shown in \Fig{f:real_spectrum}, until \Sec{s:EXP}, which addresses the experimental study of magnon BEC formation in ferrimagnetic YIG.

\subsection{\label{ss:LKI}Linear stage of the kinetic instability}
To clarify the physics of the kinetic instability, we substitute the RJ distribution\,\eqref{RJ} into \Eq{gamma} for the damping frequency. We see that near the equilibrium $\gamma_{\B k}>0$, meaning that the wave system, being close to the equilibrium, monotonically relaxes toward it. However, the right-hand-side (RHS) of \Eq{gamma} has the negative term proportional to $n_{\B 2}n_{\B 3}$, which under some conditions may dominate. 
 
To demonstrate this, let us conside the distribution as a sum of the equilibrium waves \eqref{RJ} and a  package of the parametric waves with  $k=k\sb{par}$ and total number $N\sb {par}$. In the isotropic 3D case:   
\begin{subequations}
	\begin{align}  \label{dist}
     n(k)&=n\Sp{RJ}(k)+n\sp {par}(k)\,, \\ \label{S-theoryC}
   n\sp {par}(k) &=\frac{ N\sb {par}}{4\pi k\sb {par}^2}   \delta(k-k\sb {par})     \ .
    \end{align}  
\end{subequations}

Consequently, the rate \Eqs{balance} for the bottom and top quasiparticles  $n \sb {bot}(k)$ and $n \sb {top}(k)$ appearing in the scattering process \eqref{KI-1}, are as follows:
\begin{align}\label{31} 
	\begin{split}
    \frac{\partial  n \sb {bot}(k) }{\partial t}& =  -  \gamma\sb{bot}  (k) n \sb {bot}(k)-\gamma \Sb{KI}(   k )\big [  n \sb {bot}(k)+ n \sb {top}(k)] \,, \\
     \frac{\partial  n \sb {top}(k) }{\partial t}& =  -  \gamma \sb{top} (k) n \sb {top}(k)-\gamma \Sb{KI}(   k )\big [  n \sb {bot}(k)+ n \sb {top}(k)] \ .
    \end{split}
\end{align} 
Here, $\gamma\sb{bot}>0$ and  $\gamma\sb{top}>0$ are the original (positive) damping frequencies originating from the equilibrium quasiparticles $n_{\B k} \Sp{RJ}$.   
The new terms, proportional to $\gamma\Sb{KI} (k)<0$, which leads to the kinetic instability,  can be found from the last term $-n\sb{par}(\B k_2) n\sb{par}(\B k_3)$ in the RHS of \Eq{gamma} for $^4\gamma_{\B k}$ in both the bottom and the top quasiparticles
\begin{subequations} 
	\begin{align}
 	\begin{split}\label{GammaA} 
	 &  \hskip -.1  cm \gamma \Sb{KI} (k) = - \frac{\pi}{4} \!\! \int \!\! d\textbf{\textit{k}}_1 d\textbf{\textit{k}}_2 d\textbf{\textit{k}}_3  \delta(\textbf{\textit{k}}+\textbf{\textit{k}}_1-\textbf{\textit{k}}_2-\textbf{\textit{k}}_3)\\ 
 &  \hskip -.1  cm  \times  	|W _{\textbf{\textit{k}} \B 1 }^{ \textbf{2} \textbf{3}}|^2	 \delta (\omega_\textbf{\textit{k}}  +\omega_1 -\omega_2 -\omega_3 )  
n\sp{par}(\B k_2) n\sp{par}(\B k_3) \ . 
    \end{split} 
	\end{align}   
 
As we see (following Ref.\,\cite{Lavrinenko1981}), the value of $\gamma \Sb{KI} (k)$  is negative. For  clarity of the presentation, it is convenient to introduce a positive object  $\Gamma \Sb {KI} (k) = -\gamma \Sb{KI} (k) >0 $ and to rewrite \Eqs{31} for the total number of the top and the bottom quasiparticles, $N\sb{bot}=\int n\sb{bot}(\B k)d \B k$ and $N\sb{top}=\int n\sb{bot}(\B k)d \B k$ as follows: 
	\begin{align}\label{32} 
 		\begin{split}
    \frac{\partial  N \sb {bot}  }{\partial t} = &\Gamma \Sb{KI} \big [  N \sb {bot} + N \sb {top} ]  -  \gamma\sb{bot}   N \sb {bot}  \,, \\
\frac{\partial  N \sb {top} }{\partial t} = &\Gamma \Sb{KI} \big [  N \sb {bot} + N \sb {top} ]  -  \gamma\sb{top}  N \sb {top} \ .
    	\end{split}	
	\end{align}  
Next, we integrate \Eq{GammaA} with respect to the directions of all $\B k_j$. Using the procedure of averaging as in \Refs{Dyachenko1992,Semisalov2021}, we conclude (up to a numerical prefactor) that  
	\begin{align}
 		\begin{split}\label{GammaB-2} 
  \Gamma\Sb{KI} \ (k ) &\simeq  \frac 1 k \int  d k_1 d k_2 d k_3     k_1   k_2  k_3 \, \min \{k, k_1,   k_2,  k_3\}	|W_{\B {k}    \B 1 }^{ \textbf{2} \textbf{3}}|^2	 \\ 
 \times & \delta (\omega_\textbf{\textit{k}}  +\omega_1 -\omega_2 -\omega_3 ) 
 n\sp{par}(\B k_ 2) n\sp{par}(\B k_ 3) \ .
    	\end{split} 
	\end{align}
\end{subequations}  
Note that in our case $\min \{k, k_1, k_2, k_3\}=k$ which cancels against the prefactor of the integral $1/k$.

Substituting \Eq{S-theoryC} for $n\sp{par}(\B k )$ we finally arrive at the following estimate
 for the positive contribution to the rate \Eq{32}, leading to the kinetic instability: 
\begin{align}
  \label{GbpA} \Gamma \Sb{KI} ( k ) \simeq \frac{(\Omega_{_W}\sp {par})^2 }{  \omega \sb {par}}\,, \quad  \Omega_{_W}\sp {par}  \=  |W_0  |^2	N\sb {par}
 \ .
\end{align}
Here, we assume for simplicity that $W_{\B {k} \B 1 }^{ \textbf{2} \textbf{3}}=W_0$ in agreement with \Eq{W} with $m=0$ and we approximate $\omega_0 (a\, k\sb {par})^2\simeq \omega \sb {par}$.   

The linear \Eqs{32} have exponential solutions
\begin{subequations}\label{34}
	\begin{equation}\label{34a}
N\sb{bot}(k,t)\propto \exp(\nu^{\pm}_kt)\,, \quad N\sb{top}(k,t)\propto \exp(\nu^{\pm}_kt)\,,
	\end{equation}
with
	\begin{align}
		\begin{split}\label{34b}
 \nu^{\pm}_k=& \Gamma\sb{par}-\frac12 \big(\gamma\sb{top}+\gamma\sb{bot}\big) \\ & \pm \frac12 \sqrt{  {\big(\gamma\sb{top}-\gamma\sb{bot}\big)^2+4\Gamma\Sb{KI}^2} } \ .
		\end{split}
	\end{align}  
The increment $\nu_k^+$ becomes positive if 
	\begin{equation}\label{34c}
\white{.}\hskip -.8cm 
\Gamma\Sb{KI}> \Gamma\Sb{KI}\sp{th}=\frac{\gamma\sb{top}\gamma\sb{bot}}{\gamma\sb {top}+\gamma\sb{bot}}\simeq \gamma\sb{bot} ,\ \mbox{for}\ \gamma\sb{bot}\ll \gamma\sb{top}  .
	\end{equation}
\end{subequations} 
This condition may be fulfilled for low-frequency waves near the bottom of their frequency spectra, where   $\gamma\sb {bot}( k)$ is small. If so, these waves become unstable, and their numbers $N\sb{bot}(t)$  and $N\sb{top}(t)$  are related as follows,
\begin{subequations}\label{instab}
	\begin{align}
		\begin{split}\label{instabA}
 &  N\sb {bot} \Gamma\Sb{KI}=   \frac {N\sb{top}}2\Big [ \gamma\sb {top}-\gamma \sb{bot} \\ &+ \sqrt{  {\big(\gamma\sb{top}-\gamma\sb{bot}\big)^2+4\Gamma\Sb{KI}^2} } \Big ]\propto \exp (\nu_k^+ t)\ ,   
		\end{split}
	\end{align}  
and grow exponentially until the nonlinear effects become significant. The description of these effects is the subject of the  \Sec{ss:NLT}.

Under stationary conditions, when $\nu_k^+=0$, the relationship between $N\sb{bot}$ and $N\sb{top}$ is even simpler:
	\begin{align} \label{instabB}
  N\sb {bot} \gamma\sb{bot}=  {N\sb{top}} \gamma\sb{top}\ . 
	\end{align}  
Near the threshold $ \Gamma \sb{par}=\Gamma\sb{par}\sp{th}, $ \Eq{34b} gives
	\begin{equation}\label{35c}
\nu^+_k \approx [\Gamma \Sb{KI}(k)-\Gamma\Sb{KI}\sp{th}(k)]\frac{(\gamma\sb{bot}+\gamma\sb{top})^2}{\gamma\sb{bot}^2+\gamma\sb{top}^2}+\dots
	\end{equation}
\end{subequations}

The condition  $\nu_k^+=0$  defines the threshold of the kinetic instability if one neglects the scattering of the bottom and top quasiparticles on the parametric ones, considered in \Sec{ss:scat}.  
 
Using the estimate \eqref{GbpA} for $\Gamma\sb{par}$, we find  from \Eq{34c} the critical  value $ N\sb {par} \sp{cr}$ corresponding to the threshold of the kinetic instability in which $\nu_k=0$:
\begin{equation}\label{36}
    \pi |W| N\sb {par} \sp{cr}\simeq  \sqrt{ \omega\sb {par} \,\gamma \sb {bot} \gamma \sb {top}\big /(\gamma\sb{bot}+\gamma\sb{top})}\ .
\end{equation}
Corrections to this estimate caused by the scattering of the bottom and parametric quasiparticles will be discussed in \Sec{ss:scat}.

\subsection{\label{ss:S-theory} Mean-field approximation for parametrically  excited waves and feedback limitation of the kinetic instability}
The statistical behavior of parametrically-excited waves in ferromagnets was intensively studied experimentally, theoretically and numerically since their discovery by Suhl in 1959\,\cite{Suhl1959} and by Schl\"omann in 1962\,\cite{Schlomann1962}. A relatively simple theory of this phenomenon in the mean-field approximation, called the ``$S$-theory'', was developed later by Zakharov, L'vov and Starobinets, and presented in their review \cite{Zakharov1975}. Further important achievements in this problem were summarised, for example, in the books~\cite{NSW1987,Lvov1993}. 
 
The evolution equations for the total number of parametrically-excited waves $N\sb {par}$ and their mean phase $\Psi\sb {par}$ (cf. Eqs.\,(5.4.13) of the book\,\cite{Lvov1993}) is our starting point. Here we augment them with the new term $\Gamma \sb{par} N\sb {par}$, which describes the loss of parametric waves due to their direct transfer to the bottom and the top waves by the kinetic instability described below. 

In the spherically symmetric case, these equations take the form 
\begin{subequations}\label{S-theory} 
	\begin{align}\label{S-theoryA} 
     \frac{d N\sb {par}}{dt}=& \big(h V \sin \Psi\sb {par} - \gamma\sb {par} -\Gamma \sb{par}\big)N\sb {par}   \,, \\ \label{S-theoryB}
     \frac{d \Psi\sb {par}}{dt}=& hV \cos \Psi\sb {par} + \Omega_{_S}\,, \quad  \Omega_{_S}\=  S N\sb {par}\,,  
	\end{align} 
Here, $h$ is the amplitude of the external homogeneous oscillating field, $V$ is the interaction amplitude of this field with the parametric waves, $\gamma \sb {par}$ is their damping frequency, and $S$ is the mean interaction amplitude of a pair of parametric waves with opposite wave vectors $\pm \B k\sb {par}$, with another pair $\pm \B k'\sb {par}$:  
	\begin{equation}
    S= \< W_{\B k\sb {par}, -\B k\sb {par}}^{\B k'\sb {par}, -\B k'\sb {par} }\>\ . 
	\end{equation}
\end{subequations} 

To estimate the additional damping $\Gamma \sb{par}$, consider KE\,\eqref{KE} with the collision term $^4$St$(\B k,t)$, given by \Eq{4-St}, with the resonance conditions\,\eqref{2-2}, in which we take $\B k=\B k\sb {par}$, $\B k_1=\B k'\sb {par}$, (where $|\B k\sb {par}|=|\B k'\sb {par}|=k\sb {par} $),
 $\B k_2=\B k \sb {bot}$, $\B k_3=\B k \sb {top}$  or $\B k_3=\B k \sb {bot}$,  $\B k_2=\B k \sb {top}$.
  
The last choice gives the same contribution as the previous one and can be accounted for by replacing the numerical prefactor $ \pi/ 4 \to \pi /2$. Corresponding values of the frequencies are as follows: $\omega_k=\omega_1=\omega\sb {par}$, $\omega_2=\omega\sb{bot}$, $\omega_3=\omega\sb{top}$. This way we get 
\begin{subequations}\label{bal5} 
  \begin{align} \begin{split}
     \label{bal5B} 
  \white{.}\hskip -.2 cm   \Gamma\sb{par} ( k\sb {par}) \! \simeq &    \frac{\pi}2  \!\!
    \int \!\!  d\B k _1 d \B k_2 d \B k_3 \,    \delta(\B k\sb{par}  +\B k_1 -\B k_2-\B k_3)\\  
 & \times  \delta  (2\omega\sb {par} -\omega\sb{bot}  -\omega\sb{top} ) \,   
 	|W _{\textbf{\textit{k}} \B 1 }^{ \textbf{2} \textbf{3}}|^2 \\ & \times \, n \sb {par}(\B k_1)\big[ \,n\sb {bot} (\B k_2) + \,n\sb {top} (\B k_3)\big] \ .
\end{split}  \end{align}
\end{subequations}
Note that the numerical prefactor here is twice as large as in \Eq{GammaA} for $\gamma \Sb{KI}(k)$. Estimating $\Gamma \sb{par} ( k\sb {par})$ from \Eq{bal5B}, in the same way we obtained the estimate \eqref{GbpA} for $\Gamma \Sb{KI}(k)$ from \Eq{GammaA}, we finally get: 
\begin{subequations}\label{est7}
	\begin{align}
		\begin{split}\label{est7A}
  \Gamma \sb{par}(k\sb{par}) & \simeq \frac{2|W|^2 N\sb {par}N_+}{\omega_0(ak\sb {par})^2}\simeq  \frac{2|W|^2 N\sb {par}N_+}{\omega \sb {par} }\,,\\
  \  N_+ & \=  N\sb {bot}+ N\sb{top} \,  .
		\end{split} 
	\end{align}  

Comparing this estimate with \Eq{GbpA} for $\Gamma \Sb{KI}(k)$, we see that $N\sb{par}\Gamma\sb{par}  $, the rate of dissipation of parametric waves due to the kinetic instability, is about  $2 N\sb{bot}\Gamma\Sb{KI} N_+$, the total input rate of the bottom and top quasiparticles. 
A more detailed analysis shows that this relationship is exact. Namely, the positive contribution $\Gamma\sb{par}$ to the rate \Eq{32}, leading to the kinetic instability, and the additional damping frequency $\Gamma\sb{bot}\sp{top}$ in \Eq{S-theoryA} are related as follows: 
	\begin{equation}\label{GbpB}
   \Gamma\sb{par}  N\sb{par} = 2 \Gamma\Sb{KI} N_+\ .
	\end{equation}  

Both effects are caused by the same 4-wave scattering
	\begin{equation}\label{scat}
      \omega(\B k_1)+\omega(\B k_2)\Longrightarrow \omega(\B k_3)+ \omega(\B k_4)\,,
	\end{equation}
in which the ``initial'' waves with wave vector $\B k_1$ and $\B k_2$ are parametric waves with $\omega(\B k_1)\simeq \omega(\B k_2)\simeq \omega\sb {pump}/2$, while   the ``resulting'' waves are the bottom and top  quasiparticles  with the frequencies  $ \omega\sb{bot}$ and  $\omega\sb{top} =  2\omega\sb{par}-\omega\sb{bot}$. 
\end{subequations}

From the quantum-mechanical viewpoint, one act of the scattering\,\Eq{scat} leads to the disappearing of two parametric quasiparticles and the creation of one bottom quasiparticle and one ``top'' quasiparticle with  $\omega\sb{top}$. Therefore, the corresponding damping $-\Gamma\sb{par}  N\sb {par}$ in the rate \Eq{S-theoryA} for the parametric quasiparticles must be negative. Its modulus must be positive and exactly equal to the input contributions $2\,\Gamma\ \Sb{ KI}N_+$ to the RHS of the sum of the rate \Eqs{31} for the bottom and top quasiparticles. 

Equations\,\eqref{S-theory} have the stationary solution
\begin{equation}\label{S-statA}
      (S N\sb {par})^2= (hV)^2 - (\gamma\sb {par}+\Gamma\sb {par} )^2\ . 
\end{equation}
Furthermore, \Eqs{est7A}, \eqref{GbpB} and \eqref{S-statA} allow us to find $N\sb {par}$ for a given $hV$  and $N_+$ for a given $hV$ and $N\sb {par}$:
\begin{subequations}\label{S-stat}  
   	\begin{align}
                \label{S-statB}
      & \hskip - .1cm N \sb {par}= \frac{\omega \sb {par}}{\omega \sb {par}^2 S^2+ 4 W^4 N_{+}^2}\Big [- \gamma \sb {par}W^2 N_+ \\ 
      & \hskip - .1cm +\sqrt{(hV)^2\big (\omega\sb {par}^2 S^2+ W^4 N_+^2\big ) -\gamma\sb {par} ^2 \omega\sb {par}^2 S^2 }\,\Big ]  \,,  \nonumber
	\end{align}
	\begin{align}
		\begin{split}\label{S-statC}
      & N_+ = \frac{\omega\sb {par}}{ W^4 N\sb {par}^2}  \Big [-\gamma\sb {par} W^2 N\sb {par} \\ 
      & +\sqrt{(hV W^2 N\sb {par})^2 - S^2 (W N\sb {par})^4}\Big]     \ .       
		\end{split}
    \end{align}
Using \Eq{instabB}, one easily reconstructs $N\sb{bot}$ and $N\sb{top}$ from $N_+$:
	\begin{equation}\label{45c}
N\sb{bot}= \frac{N_+ \gamma\sb{top}}{\gamma\sb{bot}+\gamma\sb{top}}\,, \quad  
N\sb{top}= \frac{N_+ \gamma\sb{bot}}{\gamma\sb{bot}+\gamma\sb{top}}\ .
	\end{equation}

Below the threshold of the kinetic instability, when $N_+=0$, \Eq{S-statB} simplifies to 
    \begin{equation}\label{S-statD}
        |S|N\sb {par}=\sqrt{(hV)^2-\gamma\sb {par}^2}\ .
    \end{equation}
    In addition, the threshold amplitude   $h\sb {th}$ of the parametric instability (for which $N\sb {par}=N_+=0$) reads
	\begin{equation}\label{S-statE}
      h \sb{th} V= \gamma \sb {par}\ .
	\end{equation} 
\end{subequations}

%%%%%%%%%%%%%%%%%%%%%%%%%%%
\subsection{\label{ss:scat} Scattering of the bottom and parametric waves }
In the nonlinear theory of kinetic instability, we have to account for one more process: scattering of the bottom  {or top quasiparticles on the intense  parametric quasiparticles with the conservation law~\eqref{2-2}, in which  $\omega_1\simeq \omega_2\simeq \omega\sb {par}$ and $\omega_{\B k}\simeq \omega_{\B 3}\simeq\omega\sb{bot}$ or $\omega_{\B k}\simeq \omega_{\B 3}\simeq\omega\sb{top}$:
\begin{equation}\label{scat2} 
\omega( k\sb{bot}) + \omega(k\sb {par})= \omega( k\sb{bot}') + \omega(k\sb {par}')\ .
\end{equation}
To do this, we have to account for an additional term St$\sb{scat}$ 
in the RHS of the rate \Eq{31}: 
 \begin{subequations}\label{KE-4}
\begin{align}\label{KE-4a}
 \frac{\partial n_{\B k}}{\partial t} = \nu_{\B k}^+ n_{\B k} +\mbox{St}\sb{scat}(\B k)\,,  
\end{align}
where $\nu_{\B k}^+$ is given, with required accuracy, by \Eq{35c} and  
\begin{align}\begin{split}
 \mbox{St} \sb{scat}(\B k)  =& \frac{\pi}{4} \int d \B k_1 d \B k_2 d\B k _3 |W_{\B k, \B k_1}^{\B k_2,\B k_3}|^2
  n\sb{par}(\B k_1)\\
  & \times   n\sb{par}(\B k_2)\delta (\omega_\textbf{\textit{k}}  +\omega_{\B 1} -\omega_{\B 2} -\omega_{\B 3} ) \\
   & \times  (n_{\B 3}-n_{\B k})  \delta(\textbf{\textit{k}}+\textbf{\textit{k}}_1-  \textbf{\textit{k}}_2-\textbf{\textit{k}}_3)\ .
\end{split}\end{align}
 \end{subequations}%\end{document}
In \Eq{KE-4a} and below in this section, $ n_{\B k}$ should be understood as $n\sb{bot}(\B k)$ or $n\sb{top}(\B k)$. This term originates from the collision term $^4$St$(\B k,t)$ [\Eq{4-St}], in which we account only for the leading terms $n\sb{par}(\B k_1) n\sb{par}(\B k_2)$ with the frequencies  $\omega_1\simeq \omega_2\simeq \omega\sb {par}$.  
 
In the isotropic case, following \Refs{Dyachenko1992,Semisalov2021}, we can rewrite \Eqs{KE-4} in the $\omega$-representation: 
 \begin{subequations}
  \label{kin-om} 
  \begin{align}
   \begin{split}
 \label{kin-omA} 
 %~\hskip - .3cm
 & \frac{\partial n_{\omega}}{\partial t}  =       
\nu_{\omega}^+ n_{\omega} +  
   \int  d\omega_1 d\omega_2 d\omega_3 \, \C   S(\omega, \omega_1, \omega_{2}, \omega_{3}) \\
   \times & \delta (\omega +\omega_{  1} -\omega_{  2} -\omega_{  3} ) \,   
 	 n_{\omega_ 1}\sp {par} n_{\omega_ 2}\sp {par}  (n_{\omega_ 3}- \sqrt{\frac{\omega_3}{  \omega}}\, n_{\omega}) . \end{split}\end{align}  
Here $\omega$ and $k$ are related by \Eq{3}: $\omega=\omega_0[1+(ak)^2]$, and the particle number densities in the $k$- and $\omega$-spaces, $n_k$ and $n_\omega$, in isotropic  case are related as follows,
\begin{equation}
     \label{rel-o-k}
     n_\omega=\frac{2\pi k }{\omega_0a^2}\,n_k\ .
\end{equation}
It is important  to take into account that parametrically excited waves usually  
experience auto-oscillation with a characteristic frequency about   $\Omega\Sb S$, given by \Eq{S-theoryB}    \,\cite{Lvov1993}.
Therefore, in the $\omega$-representation the distribution $n\sb p(\omega)$ is not proportional to $\delta (\omega-\omega\sb p)$, but has some width $\Delta$ of the order of $\Omega \Sb{S} $ around $\omega\sb {par}$. For concreteness,  we assume the simple Gaussian form of $n\sp {par}(\omega)$: 
\begin{align}\begin{split}
      \label{bal41}
  n\sp{par}_\omega= &\frac {N \sb {par}}{\sqrt{2\pi}\Delta}\exp\Big [-\frac{(\omega-\omega\sb {par})^2}{2\Delta^2}\Big]\,,\\
  N\sb {par}= &\int\limits _{-\infty}^\infty n\sp {par}_\omega d \omega \,, \quad \Delta \simeq \Omega\Sb S \simeq S N\sb{par} \ . 
 \end{split}   
\end{align} 
\end{subequations} 
   
The collision integral in the RHS of \Eq{kin-omA} is taken over the positive values $\omega_1$, $\omega_2$, and  $\omega_3$. The kernel of the integral (up to a dimensionless order-one constant) reads:
\begin{eqnarray} \label{kin-omB} 
\white{.}\hskip - .8cm \C S(\omega, \omega_1, \omega_{2}, \omega_{3})\simeq   \frac{\min\{ \sqrt {\omega},  \sqrt {\omega_1}, \sqrt {\omega_2}, \sqrt {\omega_3}\} }{ \sqrt{ \omega_1 \omega_2 \omega_3}}|W_{ k,   1 }^{  {2}  {3}}|^2   .
\end{eqnarray} 
  Substituting  $n\sp{par}_{\omega_1}$ and   $n\sp{par}_{\omega_2}$ from \Eq{bal41} and keeping in mind that in this case $\omega\simeq \omega _3\simeq \omega_0< \omega_1\simeq \omega_2\simeq \omega\sb{par}$ we conclude that 
\begin{equation} \label{kin-omD}
\C S(\omega, \omega_1, \omega_{2}, \omega_{3})\simeq \frac{|\C W|^2}{\omega\sb{par} }\,, \quad
  \C W=  W_{ k_0,   k\sb {par}}^{ k\sb {par} k_0,} \ .
\end{equation}  
 
Furthermore, replacing the dummy variable $\omega_3$ by $\widetilde \omega$ and integrating the resulting equation over $\omega_1$ and $\omega_2$, we get (up to a numerical prefactor in the integral)
\begin{align} 
 \label{fin-balA} \nonumber
 \frac{\partial n_{\omega}}{\partial t} & =      
  \nu_{\omega}^+ n_{\omega} +  
  \frac{\Omega_{_W}^2 }{\omega\sb {par}}\int \limits _{\omega_0}^\infty \frac{d\widetilde \omega}{ \Delta   } 
       (n_{\widetilde \omega }-n_{\omega})   \exp\Big [-\frac{(\omega-\widetilde \omega )^2}{4\Delta^2}\Big]\,, \\
       \Omega\Sb{W}& = \C W N\sb {par} \ .
\end{align} 

Equation \eqref{fin-balA} was derived for the 3D case. In 2D, it has exactly the same form\,\eqref{fin-balA} but with a slightly different value of the numerical prefactor, which we are not controlling anyway.
   
Already at this stage, we can formulate an important consequence of \Eq{fin-balA}: integrating it over $\omega$ from $\omega_0$ to $\infty$ one gets a rate equation for the total number of bottom quasiparticles $N\sb {bot}$ (or $N\sb{top}$):
\begin{equation}\label{Np-int}
     \frac{d N\sb {bot}}{d t}= \< \nu_\omega ^+\> N\sb {bot} \,, \quad \< \nu_\omega ^+\>\= \frac{\int_{\omega_0}^\infty \nu_\omega^+  n_\omega d\omega}{\int_{\omega_0}^\infty n_\omega d\omega}\ .
\end{equation}
The integral term in \Eq{fin-balA} does not contribute to \Eq{Np-int} due to the anti-symmetry of its integrand.

\subsection{\label{ss:NLT}Nonlinear theory of the kinetic instability}
Some aspects of the nonlinear theory of the kinetic instability were discussed a long time ago in  \REF{Lvov1981c} focusing only on the case of small super-criticality over the threshold of kinetic instability when $\Gamma\Sb{KI}-\Gamma\sp{th}\Sb{KI} \ll \Gamma\sp{th}\Sb{KI}$. Here, we extend this theory to the range of moderate and large super-criticality, i.e., when $\Gamma\Sb{KI}-\Gamma\sp{th}\Sb{KI} \gtrsim \Gamma\sp{th}\Sb{KI}$ or when $\Gamma\Sb{KI}\gg \Gamma\sp{th}\Sb{KI}$. 
 
The goal of the nonlinear theory of kinetic instability, developed here, is to find the spectral profile of the bottom quasiparticles $n_\omega$, the total number of the parametric, bottom, and top quasiparticles, $N\sb {par}$, $N\sb {bot}$, and $N\sb{top}$, as a function of the pumping amplitude $h$, accounting for the main interactions in the system only. The interactions among the parametric waves in the $S$-theory approximation are described in \Sec{ss:S-theory}, the particle number flux from the parametric to the bottom quasiparticles -- in \Sec{ss:LKI} and scattering of the bottom and parametric quasiparticles -- in \Sec{ss:scat}. At this stage, we neglect other nonlinear effects that might be important depending on the particular characteristics of the system at hand (e.g., the value and orientation of the external magnetic field, etc). The list of possibly important effects includes cascade mechanisms of   particle transfer from  the parametric and top quasiparticles to bottom ones, as described in \Sec{s:FE-SP}, and nonlinear   interactions in the  system of the bottom quasiparticles, which can lead to the redistribution of the bottom quasiparticles and the growth of the damping frequency of the bottom quasiparticles with their number (see, e.g., \REF{Lvov1982b} and book\,\cite{Lvov1993}). 
  
\subsubsection{Narrow  package approximation}

\begin{figure} [b]
	\includegraphics[width=9cm]{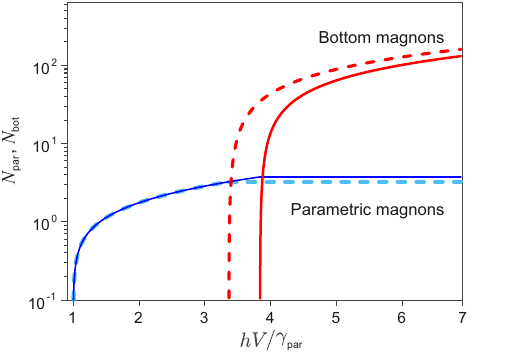} 
	\caption{ \label{F:3}  Qualitative representation of the normalized number of parametric $S N\sb {par}/\gamma\sb {par}$ (light blue dashed line)  and bottom quasiparticles $S N\sb {bot}/\gamma\sb {bot}$ (red dashed line)  according to \Eqs{S-stat}. For concreteness, we took $S=W$, $\gamma\sb {par}=\gamma\sb {bot}\ll \gamma\sb{top}$, $\omega\sb {par}=1000 \gamma\sb {par}$. Below the threshold of kinetic instability, $S N\sb {par}/\gamma\sb {par}$ is defined by \Eq{S-statD}. Above the threshold, $S N\sb {par}/\gamma\sb {par}$ is frozen at its value at the threshold defined by \Eq{36}. $N\sb{bot}$ is a fraction of $N_+$ according to \Eq{45c}. In its turn, $N_+$ as a function of $h$ is defined by \Eq{S-statC} with $ N\sb {par}=N\sb {par}\sp{cr}$. Note that $N\sb{bot}=0$ for $h\le h\sb{cr}$. 
		Solid lines: the same, but accounting for the scattering of the bottom quasiparticles on the parametric ones in the framework of the exponential model  (cf. \Sec{sss:mom}). Here the spectral width of package  is $\C D\sb {par}=15$, giving $\lambda _0\sp{mod}=0.26$.  
	}
\end{figure}

Assuming  initially  that the package $n_\omega$ is extremely narrow, such that $\nu_\omega$ can be considered a constant, denoted $\nu_0=\nu^+_{k_0}$, one obtains from \Eq{fin-balA}:
\begin{align} 
            \label{Np-intB}
     \frac{d N\sb {bot}}{d t}& =  \nu_0   N\sb {bot}   \, .
\end{align}

Below the kinetic instability threshold, when $\nu_0< 0$, the total number of the bottom quasiparticles decays exponentially, and in the stationary condition $N\sb {bot}=0$ holds. Thus, according to \Eq{S-statD}, $N\sb {par}$ increases with the pumping amplitude $hV$, see \Fig{F:3}, dashed light-blue line, until it reaches the value $N\sb{par}\sp{cr}$ given by \Eq{36}.  

The balance between the number of bottom and parametric quasiparticles is maintained by the increments $\nu^+$ and  $\Gamma\sb{par}$. As $N\sb {par}>N\sb {par}\sp{cr}$, the increment $\nu_{ k}^+$ becomes positive in a narrow range around $k_0$, and $N\sb {bot}$ starts growing exponentially according to \Eq{instabA}. The bottom quasiparticles take energy from the parametric ones as described by the additional damping frequency $\Gamma\sb{par}$ in the rate \Eq{S-theoryA} for the number of parametric quasiparticles $N\sb {par}$. As a result, $N\sb{par}$ drops back to $N\sb {par}\sp{cr}$ and the increment $\nu_{k_0}^+ \to 0$. Therefore, $N\sb {par}$ becomes frozen at the level $N\sb {par}\sp{cr}$ for any $h>h\sp {cr}$, see the horizontal  dashed line in \Fig{F:3}. 

Substituting in \Eq{S-statC} $N\sb{par}=N\sb{par}\sp{cr}$ from \Eq{36} and using \Eq{S-statD} for $h=h\sb{cr}$, we obtain an equation for $N_+$ in which we assume for simplicity that $W=S>0$
\begin{subequations}\label{simple}
	\begin{equation}\label{simpleA}
 SN_+=\frac{\omega\sb{par}}{S N\sb{par}\sp{cr}}\Big[-\gamma\sb{par}+ \sqrt{\gamma\sb{par}^2 +V^2(h^2-h\sb{cr}^2)} \,\Big]\ .
 	\end{equation}
 For small super-criticality over the threshold of the kinetic instability $\delta h\= h- h\sb{cr}\ll h\sb{cr}$, one gets from \Eq{simpleA}
	\begin{equation}\label{simpleB}
 SN_+\simeq SN\sb{cr}\,  \frac{\omega\sb{par}}{\gamma\sb{par}} \, \frac {\delta h}{h\sb{cr}} \ .
	\end{equation}
\end{subequations}
One sees from \Fig{F:3} that $N\sb {bot}$ grows sharply just above $h\sp{cr}$ as predicted by \Eq{simpleB}:
$N\sb {bot}$ reaches the level of $N\sb {par}\sp{cr}$ (crossing of the blue and red lines) for very small $(\delta h/h \sb{cr})\simeq \gamma \sb{par}/\omega\sb{par}\ll 1$. 

Now, we include the scattering of the bottom quasiparticles on the parametric ones, as described by the integral term in \Eq{fin-balA}. Then  \Eq{Np-int} is identical to \Eq{Np-intB} upon  replacement  of  $\nu_{ 0}$ by the mean value $\< \nu_\omega\>$.
Due to the linearity of scattering  \Eq{fin-balA}, the profile of the bottom quasiparticles $n_\omega$ is independent of their total number $N\sb {bot}$. Instead, it depends only on the number of the parametric quasiparticles $N\sb {par}$, which is constant for $h>h\sp{cr}$. 
Therefore, in the estimation of $\Omega_{_W}=\C W N\sb{par}$ and $\Delta\simeq |S|N\sb {par}$ in \Eqs {fin-balA} and \eqref{bal41} we have to take $N\sb {par}=N\sb {par}\sp {cr}$. Thus, we conclude that $n_\omega$ and $\< \nu_\omega\>$ are $h$-independent for $h>h\sp{cr}$. 
  
\subsubsection{\label{sss:num}Numerical analysis of the particle rate equation }
To extend the nonlinear theory of the kinetic instability beyond the narrow package approximation, we studied the rate \Eq{fin-balA} numerically. 
It is convenient to do this by transforming \Eq{fin-balA} to the dimensionless form by introducing $\tau= \gamma \sb {bot} t$, $\C D \sb {par} =\Delta/\gamma\sb {bot}$, $x=(\omega-\omega_0)/\gamma\sb {bot}$ and $y=(\widetilde \omega -\omega_0)/\gamma\sb {bot}$:
\begin{subequations}\label{ND}
 \begin{align} \label{ND-A}
   \begin{split}
  ~\hskip - .3cm \frac{\partial n_{x}}{\partial \tau}  =&      
\lambda _{x} n_{x} + \mbox{Int}_x\,,\\  \mbox{Int}_x \=& \sqrt{\frac 2 \pi}  
  \frac{\C A} {\C D\sb {par}}  \int \limits _0^\infty  dy  
       (n_{y}-n_{x}) \exp\Big [-\frac{(x-y)^2}{2\C D\sb {par}^2}\Big]  \ .
   \end{split}
 \end{align}  
Here, the dimensionless increment is $\lambda_x=\nu_\omega/\gamma\sb {bot}(\omega_0)$. Assuming for concreteness
  $\gamma\sb {bot}(\omega) \propto \omega$, i.e. $ \gamma\sb {bot}(\omega)=\gamma\sb {bot}(\omega_0) \omega/\omega_0$, we have
 \begin{align}\label{ND-B}
      \begin{split}
                \lambda_x=& \lambda  _0 - \frac{x}{x_0}\,, \quad  \lambda_0 \=\Big ( \frac{N\sb {par}}{N\sb {par}\sp {cr}}\Big )^2-1 \,, \\ x_0=& \frac{\omega_0}{\gamma\sb {bot}(\omega_0)}\ .
   \end{split}
 \end{align}
  Taking into account \Eqs{fin-balA}, we estimate  
 \begin{equation}\label{ND-C}
      \C A \simeq \frac {\omega\sb {par} }{4\pi\omega _0}\Big ( \frac W {\widetilde W} \Big )^2 
 \end{equation}
\end{subequations}
which is of the order of unity. 
  
To get more detailed information about $n_x$ and  $\lambda_0$, we numerically solve \Eq{ND-A} together with the $S$-theory \Eqs{S-theory} for $n_\omega$ using $\C D\sb {par}=5, 15$ and $45$, and taking for concreteness $x_0=100$. Resulting profiles of $n_x$ are shown in \Fig{f:3} by color solid lines together with their approximate exponential fits,
\begin{equation}\label{exp}
      n_x=\frac 1 {\C D\sb {bot} }\exp\big ( - \frac x {\C D\sb {bot} }\Big ),
\end{equation}
shown by dashed lines with matching colors. The values of $\lambda_0=\lambda_0\sp {num}$, corresponding to the stationary conditions, and $\C D \sb {bot}=\C D \sb {bot}\sp {fit}$ for these three values of $\C D\sb {par}$ are given in the Tab.\,\ref{t:1}. 
  
\begin{table} [b]
      \centering
      \begin{tabular}{c| c||c| c | c  }
     1~~&  $\C D\sb {par}$ & ~~~5~~~ & ~~~15~~~ & ~~~45~~~   \\ \hline \hline
     2~~&   $\< \lambda  _x\> \sp {num}$, \Eq{ND}   &  \, $0.13\pm 0.01$   & \, $0.27\pm 0.01$    &   $0.46\pm 0.01$    \\ 
     3~~&    $\< \lambda  _x\> \sp {mod}$, \Eq{Db-Dp}  &   0.13  & 0.26  & 0.48   \\ 
     \hline
     4~~&  $\C D\sb {bot}\sp {fit}$, \, \Eq{exp}  &~   $19\pm 5$ ~ &~ $33\pm 5$~  &~ $46\pm 5$   ~
        \\
       5~~&  $\C D\sb {bot}\sp {mod}$, \Eq{Db-Dp}   &   13 & 26 & 48 
        \\  
            6~~&  $\C D\sb {bot}\sp {mix}$, \Eq{zero-momB}   &   13 & 27 & 48 
        \\    
           
      \end{tabular}
      \caption{\label{t:1}  The increments $\lambda_x$ and  the width $\C D\sb {bot}$ for three values of $\C D\sb {par}$. The quantities are listed in lines: (1): $\C D\sb {par}$ [cf. \Eqs{ND}]; (2): Numerically found value of $\< \lambda  _x\>\sp{st}=\< \lambda  _x\>\sp{num}$, \Eqs{ND}; %for different value of $\C D\sb {par}$  which provides stationary  solution of \Eqs{ND}. 
      (3): $ \< \lambda  _x\>\sp{st}=\< \lambda  _x\>\sp{mod}$, found from the exponential model \Eq{zero-momB};
     (4): The approximate width $\C D\sb {bot}\sp{fit}$ of exponential fit \Eq{exp}, shown in \Fig{f:3} by dashed lines. 
     (5):$\C D \sp{mod}$; 
     (6): $\C D\sb b\sp{mix}$ -- ``mixed'' value of $\C D\sb {bot}$ found by the model \Eq{zero-momB}  in which  $\lambda  _0$ is numerically found $ \<\lambda  _x\> \sp {num}$ shown in the   line 2. } 
\end{table}

\begin{figure}  [t] 
    \includegraphics[width=9.5cm]{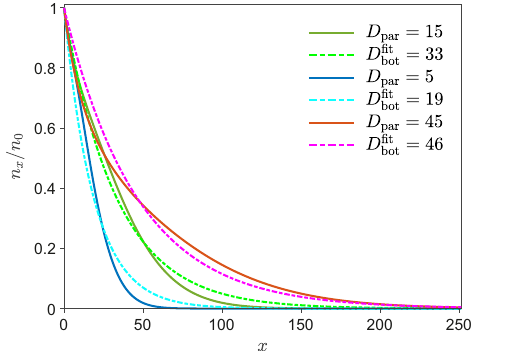}
  \caption{ \label{f:3} 
  Normalized stationary solutions of \Eqs{ND-A} $n_x/n_0$ for $\C D\sb {par}=5$, (red line) $\C D\sb {par}=15$, (blue line)   and 
 $\C D\sb {par}=45$ (pink line) with $x_0=100$ and $\C A=1$. Respectively colored dashed lines show their approximated exponential fit with \Eq{exp}. 
 } 
\end{figure}   
       
\subsubsection{\label{sss:mom}Exponential model of the bottom  quasiparticles  distribution} 
Based on the results of the numerical solution of \Eq{ND-A}, we assume that the profile $n_x$ has an exponential form\,\eqref{exp} with some yet unknown value $\C D\sb {bot}$. 
Under this assumption,  we have two free parameters, $\lambda _0$ and $\C D\sb {bot}$. 
To find them, we need two relations. The first relation between them comes from the rate equation for the number of bottom quasiparticles $N\sb {bot}=\int_0^{\infty} n_x dx$. We integrate the stationary \Eq{ND-A}  with $n_x$ given by \Eq{exp} and take into account that the term ${\rm Int}_x$ vanishes upon integration. Then we have:
\begin{subequations}\label{zero-mom}
	\begin{align} \label{zero-mom0}
  \frac{d N\sb {bot}}{d t}= &  \< \lambda  _x\> N\sb {bot}\,, \\
  \label{zero-momA}
      \< \lambda  _x\> \equiv & \int\limits_0^{\infty}  \lambda  _x n_x\, dx =  \lambda  _0- \frac {\C D\sb {bot}}{x_0}\ . 
    \end{align}
 Obviously,   \Eq{zero-mom0}    describes a steady state if $\< \lambda  _x\>=0$, i.e.
	\begin{equation}\label{zero-momB}
  \C D\sb {bot}= x_0  \,  \lambda  _0 \ .
	\end{equation}
\end{subequations}
The second relation between $\lambda  \sb {par}$ and $\C D\sb {bot}$ is provided by the first moment of \Eq{ND-A}. Multiplying this equation by $x$ and integrating over $x$ from zero to infinity, we find: 
\begin{subequations}\label{one-mom}
	\begin{align}\label{one-momA} 
    0&= \<x\,  \lambda  _x\> + \< x \, \mbox{Int}_x\> \,, \\
    	\begin{split} \label{one-momB}
          \<x\,  \lambda  _x\> & = \int\limits _0^{\infty}x \lambda  _x n_x\, dx = - x_0 \lambda  _0^2= -\frac{\C D\sb {bot} ^2}{x_0}\ .  
    	\end{split}
	\end{align}
\end{subequations}
To simplifyn \Eq{one-momB} we used \Eq{zero-momB}.  

With $n_x$ given by \Eq{exp}, the integral\,\eqref{ND-A} for Int$_x$ can be found analytically:
\begin{align}\label{Int} 
    \begin{split}
           \mbox{Int}_x=& n_x \Big[ 1+  \mbox{erf}  \Big ( \frac x {\sqrt 2 \C D\sb {par}}\Big )\\ & -  \exp\Big (\frac {\C D \sb {par}^2}{2 \C D\sb {bot} ^2} \Big ) \mbox{erfc}  \Big ( \frac {\C D \sb {par}^2- \C D\sb {bot} x} {\sqrt 2 \C D \sb {bot}\C D\sb {par}}\Big )\Big ]\ .
    \end{split}
\end{align}
Here, $\mathrm{erf}(z)= \frac 2 {\sqrt \pi}\int_0^z \exp (-t^2)dt$ is the Gauss error function and   $\mathrm{erfc}(z)=1-\mathrm{erf}(z)$.

Unfortunately, we cannot find the integral
  $ \< x \mbox{Int}_x\>=\int \limits_0^\infty x \mbox{Int}_x dx$ analytically. Therefore, we have solved \Eqs{one-mom} for $\C D\sb {bot}$,
\begin{align}\label{Db-Dp}
   \begin{split}   
    \C D\sb {bot}^2=& x_0 \int \limits_0^\infty x\, n_x \Big[1+ \mbox{erf} \Big (\frac x {\sqrt 2 \C D\sb {par}}\Big )\\ & - \exp\Big (\frac {\C D \sb {par}^2}{2 \C D\sb {bot} ^2} \Big ) \mbox{erfc}  \Big ( \frac {\C D \sb {par}^2- \C D\sb {bot} x} {\sqrt 2 \C D \sb {bot}\C D\sb {par}}\Big )\Big ] dx,
   \end{split}  
\end{align}
numerically. The results for $\C D \sb {bot}$ denoted $\C D \sb {bot}\sp{mod}$  with $x_0=100$ (blue line) and $x_0=300$ (red line) are shown in \Fig{f:4}b.

\begin{figure} [t] 
	\includegraphics[width=1\columnwidth]{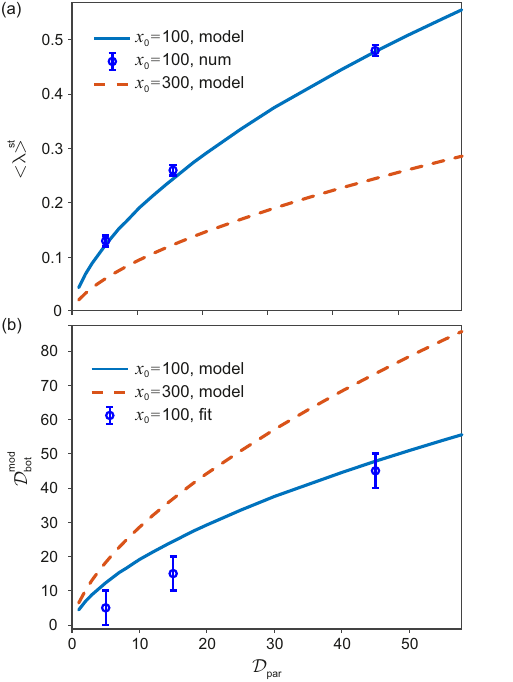}
	\caption{ \label{f:4} Stationary solutions of the exponential model for (a) $\< \lambda_x\> =\< \lambda_x\>\sp{st} $ [\Eq{zero-momA}], and (b) $\C D\sb{bot}\sp {mod}$, as a function of $\C D\sb{par}$.
	The results for $x_0=100$ are shown by solid blue lines and for $x_0=300$ by dashed red lines. Blue dots with error bars denote numerical values  $\<\lambda _x\> \sp {st}=\<\lambda_x\>\sp{num}$ [\Eq{ND-B}] for $x_0=100$.}
\end{figure}

\subsubsection{\label{sss:comp} Comparison of the numerical solution of quasiparticle rate equation and predictions of the exponential model }
  
To clarify how the solution of \Eq{Db-Dp}, derived in the framework of the approximations of the exponential model for the bottom quasiparticle distributions, corresponds to the ``exact'' numerical solution of the basic \Eqs{ND}, we compare corresponding results for the effective increment $\<\lambda_x\>$, the characteristic width $\C D\sb {bot}$ and the effective increment $\lambda_0$ obtained in both ways. 

\paragraph{Effective increment $\<\lambda_x\>$.} The values of $\<\lambda_x\>\sp{st}=\<\lambda_x\>\sp{num}$,  found from numerical  stationary  solution of \Eqs{ND}, are given in line 2 of Tab.~\ref{t:1}. Corresponding values of $\<\lambda_x\> \sp{mod}$, obtained from the exponential model \Eq{zero-momB} in which $ \C D \sp{mod}\sb {bot}$ is the solution of the model \Eq{Db-Dp}, are listed in line 3. To get a more general view on the model dependence $\<\lambda_x\>$ vs $\C D\sb {par}$, we have presented these dependencies in \Fig{f:4}(a)  for $x_0=100$, blue line, and $x_0=300$, red line. Blue dots with error bars denote values of $\<\lambda_x\>\sp{num}$ found numerically for $x_0=100$, $\C D\sb {par}=5,\ 15$ and $\C D\sb {par}=45$. We see a very good quantitative agreement between these two approaches.

\paragraph{Effective width $\C D\sb {\rm bot}$ in the approximate exponential distribution \eqref{exp}.} The numerical profiles, shown by the solid lines in \Fig{f:3}, where fitted by the exponential function\,\eqref{exp} by finding the parameter $\C D\sb {bot}$ which minimizes the mean-square deviation. The resulting values of $\C D\sb {bot}=\C D\sb {bot}\sp{fit}$ are given by line 4 in Tab.~\ref{t:1}, and the corresponding exponential profiles are shown in \Fig{f:3} by dashed lines. We see that although the fitted profiles describe the ``exact'' numerical profiles reasonably well, there are some systematic deviations between them. For example, for $\C D\sb {par}=45$, the fitted profile goes above the numerical one for $n_x>0.35$ while for $n_x<0.35$, the fitted profile goes slightly below the numerical profile. This means that for $n_x>0.35$ the current value of $\C D \sb {bot}<\C D\sb {bot}\sp{fit}=46 $,  while for $n_x<0.35$ the current value of $\C D \sb {bot}>\C D\sb {bot}\sp{fit}=46$. Analyzing the current values of $\C D\sb {bot}$ for different $x$, we estimate the error bars as $\pm 5$. For completeness, in line 6 of Tab.\,\ref{t:1} we also list a ``mixed'' value of $\C D\sb {bot}=\C D\sp{mix}\sb {bot} $ obtained from the model \Eq{zero-momB} in which $\lambda_0=\lambda_0\sp {num}$. To complete the comparison, in \Fig{f:4}, we have plotted the ``fit'' values of $\C D\sb {bot}=\C D\sb {bot}\sp{fit}$ (blue dots with error bars), where the model dependence $\C D\sb {bot}$ vs. $\C D\sb {par}$ is shown for the same value of $x_0=100$ by the solid blue line.  

In all the cases, we see a very reasonable agreement between the model values of $\<\lambda_x\>$ and $\C D\sb {par}$ with the corresponding ``exact'' values found from the numerical solutions of the basic \Eqs{ND}.  
 
As we have discussed above, the scattering of the bottom quasiparticles on the parametric ones results only in replacing of $\nu_0$ by $\< \nu_\omega\>$ in the rate \Eq{Np-int} for $N\sb {bot}$. In dimensionless units we need to replace $\lambda_0$, given \Eq{ND-B}, by $\< \lambda_x\>$, given in our exponential model by \Eq{zero-momA}:
\begin{equation}\label{condA}
     \< \lambda_x\>= \Big(\frac{N\sb {par}}{N\sb {par}\sp {cr}}\Big)^2- 1 - \frac{\C D\sb {bot}}{x_0}\ .
\end{equation}
As before, the stationarity of the bottom quasiparticles requires $  \< \lambda _x\>=0$, which can be achieved in the presence of scattering for some $N\sb {par}\= N\sb {par} \sp{st}> N \sb {par}\sp {cr}$. Denoting the stationary value of $ \< \mu _x\>$ as  $ \< \mu _x\>\sp {st}$, we can write
\begin{equation}\label{condAa}
     \< \mu _x\>\sp{st}= \Big(\frac{N\sb {par}\sp {st}}{N\sb {par}\sp {cr}}\Big)^2- 1 - \frac{\C D\sb {bot}}{x_0}=0\,,
\end{equation}
or 
\begin{equation}\label{condB}
    N\sb {par}\sp {st}=N\sb {par}\sp {cr} \sqrt{ 1 +  \C D\sb {bot}/x_0}\ . 
\end{equation}
Now, in order to account for the scattering of the bottom quasiparticles, we need to replace $N\sb p\sp{cr}$ by $N\sb {par}\sp{st}>N\sb {par}\sp{cr}$ in \Eqs{S-statC} and \eqref{S-statD} for the dependence of $N\sb {par}$ and $N\sb {bot}$ on $hV$. 
Considering for concreteness the case $\C D\sb {par}=15$, we find $\lambda_0\sp {mod}=0.26$; see Tab.\,\ref{t:1}. In \Fig{F:3} (solid lines), we have plotted the total number of the parametric $N\sb{par}$ and the bottom quasiparticles $N\sb {bot}$ vs. amplitude of the parametric pumping $hV$ taking for concreteness $h\sp {cr}=10 \gamma\sb {par}$.  

We see that for large enough $hV$ the number of the bottom magnons can essentially exceed the number of the parametric magnons. In this case, besides the scattering of the bottom magnons on the parametric ones, one should also account for the four-magnon scattering in the subsystem of the bottom magnons and for the Kolmogorov-Zakharov cascade of the top magnons down to the bottom ones. The corresponding theory is beyond the scope of the present paper.     

%%%%%%%%%%%%%%%%%%%%%%%%%%%%%%%%%
  
\section{\label{s:EXP}Experimental results, discussion and comparison with theory} 
In the preceding sections,  we discussed possible regimes of Bose-Einstein condensation for  various intensities of the energy input. In \Sec{s:FE-WP}, we studied the BEC scenario under weak pumping, when quasiparticles ( e.g.,  magnons) near the bottom of their frequency spectrum are in local thermodynamic equilibrium. In the following \Sec{s:FE-SP}, we addressed Bose-Einstein condensation in the case of strong pumping, when quasiparticles are transferred from the pumping region to the BEC region at the bottom of the frequency spectrum by a step-by-step cascading Kolmogorov-Zakharov process. Section\,\ref{s:KI} was focused on the situation of ultra-strong pumping, in which there is a direct transfer of the pumped magnons to the spectrum bottom due to the process of kinetic instability. In particular, in \Sec{ss:NLT}, we developed a nonlinear theory of the kinetic instability that allowed us to find the frequency distribution of the bottom magnons and to estimate the number of magnons in both the pump and BEC regions as a function of pumping power.

The kinetic instability regime considered in \Sec{s:KI} appears to be the most interesting from both theoretical and practical points of view. Being the physically most nontrivial regime, it also allows the formation of the densest magnon condensates suitable for practical applications. 
Therefore, in this \Sec{s:EXP}, devoted to the experimental results, we focus on the case of ultra-strong parametric pumping of magnons. Here, we will compare our theoretical conclusions with both existing and new experimental data obtained in our work. Our main goal is to determine the main factors contributing to the transition of magnons toward the lower part of their frequency spectrum in YIG films and to analyze their frequency distribution in the BEC region.

%%%%%%%%%%%%%%%%%%%%
\begin{figure}[b]
	\includegraphics[width=1\columnwidth]{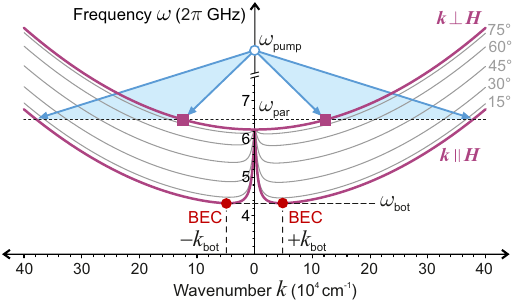}
    \caption{
		\label{f:real_spectrum}
		The spectrum of the fundamental magnon mode in a 5.6-\textmu m-thick YIG film planarly magnetized by a magnetic field $H=1500$\,Oe. The spectrum is shown for the wave vector $\bm{k} \parallel \bm{H}$, for $\bm{k} \perp \bm{H}$, and for several intermediate wave vector directions (gray curves). The blue arrows and light blue shadow areas illustrate the process of magnon injection by parallel parametric pumping with frequency $\omega_\mathrm{pump}$. The frequency of parametric magnons $\omega_\mathrm{par}=\omega_\mathrm{pump}/2$ is marked with a dotted line. The red dots indicate the positions of the frequency minima $\omega\sb{bot}(+\bm{k}\sb{bot})$ and $\omega\sb{bot}(-\bm{k}\sb{bot}$) occupied by $+k$- and $-k$-BECs of magnons. The two magenta squares show the magnon pair with the lowest threshold of parametric instability \cite{Serga2012}. 
            }    
\end{figure}
%%%%%%%%%%%%%%%%%%%%

\subsection{\label{ss:spectrum} Actual frequency spectrum in YIG films} 

Most experimental studies of the magnon Bose-Einstein condensation, as in this work,  have been carried out in tangentially magnetized YIG films using parametric pumping and Brillouin light scattering (BLS) spectroscopy. This is because microwave parametric pumping is one of the most efficient methods of magnon injection, and BLS spectroscopy allows access to the broad frequency--wave-vector domain of the magnon spectrum. The strong reduction of the BLS signal upon transition to normal magnetization of a magnetic film favors the use of tangential magnetization geometry. 
The low-damping ferrimagnetic YIG films provide the highest possible ratio between the spin-lattice relaxation time of the magnon gas and the thermalization time of the pumped magnons, which motivates the preference of these films over metallic ferromagnetic films or Heusler compounds. 

To start our analysis of the experiment, we performed numerical calculations of the magnon frequency spectrum corresponding to our experimental conditions, i.e., in a 5.6-\textmu m-thick YIG film tangentially magnetized by the magnetic field $H=1500$\,Oe (see Fig.\,\ref{f:real_spectrum}). 
This spectrum qualitatively corresponds to the spectra of all magnetic films with thicknesses ranging from a few microns to tens of microns, widely used in experimental studies of kinetic instability and Bose-Einstein condensation of magnons. In an unbounded film, the spectrum of magnon modes is discrete in the direction $z$ normal to the film plane and continuous in its plane. Magnons condense at the lowest-frequency (fundamental) mode with the homogeneous or quasi-homogeneous distribution of dynamical magnetization over the film thickness.
This mode $\omega(k_\|,k_{\perp},k_z=0)$ is shown in Fig.\,\ref{f:real_spectrum}. 
 
Unlike the spectrum presented in Fig.\,\ref{f:0}, the spectrum of magnons in tangentially magnetized films is strongly anisotropic, being very different for $\B k = \B k_\| \parallel \bm{H}$ (lower part of the spectrum, magenta curves), and for $\B k = \B k_\perp \perp \bm{H}$ (upper part, magenta curves). Gray curves show several intermediate directions of in-plane wave vectors $\B k=(k_\|,k_{\perp}, k_z=0)$.
It is remarkable that for $\B k_\| \parallel \bm{H}$ the spectrum has two equivalent minima with $k_\|=\pm k\sb{bot}$ (with $k\sb{bot}\approx 4.5\cdot 10^4\,$cm$^{-1}$). 
The red dots indicate the positions of these frequency minima $\omega\sb{bot}(+\bm{q}\sb{bot})$ and $\omega\sb{bot}(-\bm{k}\sb{bot}$) occupied by $+k$- and $-k$-BECs of magnons.

Blue arrows and light blue shadow areas illustrate the process of magnon injection by parallel parametric pumping leading to filling by parametric magnons the entire surface $\omega (\B k)=\omega\sb{pump}/2$ for large pumping power\,\cite{Lvov1993}.  

Despite the significant difference between the spectra presented in Fig.\,\ref{f:0} and Fig.\,\ref{f:real_spectrum}, the qualitative results obtained in the previous sections remain valid. Both cases can be described by the same Gross-Pitaevskii equations with the position of the frequency minima properly rescaled, as well as the scale of wave vectors in their vicinity.

\subsection{\label{ss:exp}Experimental procedure} 
The experiments were performed using samples with a YIG film of a thickness 5.6\,{\textmu m} and 6.7\,{\textmu m}. All samples were grown by liquid-phase epitaxy in (111) crystallographic plane on a gallium gadolinium garnet substrate. 

To detect the magnons, we used Brillouin light scattering spectroscopy that allowed us to obtain the frequency-, wave-vector-, space-, and time-resolved spectra. It was equipped with electromagnetic parametric pumping circuits. 

\begin{figure} 
\includegraphics[width=1\columnwidth]{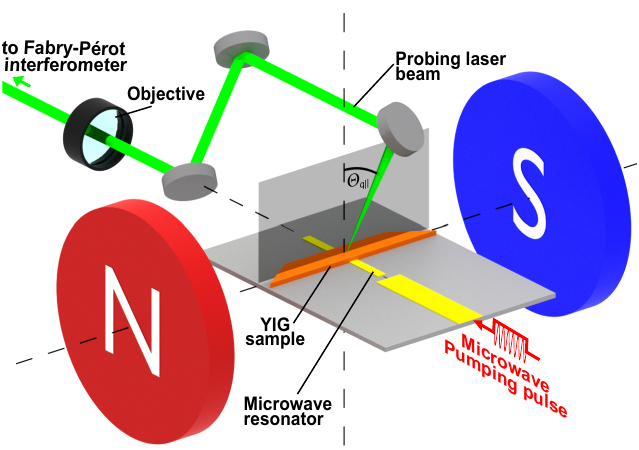}  
   \caption{\label{f:2} Schematic setup of the wave-vector-resolved BLS experiment. The probing laser beam is focused on the parametrically pumped area of the YIG sample by an objective lens. The laser light incidence angle $\Theta_{q \parallel}$ is steered using a combination of three dielectric mirrors mounted on a rotary stage (not shown), allowing for a change in the incident angle from $-90^\circ$ to $90^\circ$. The plane of incidence is oriented along the direction of the bias magnetic field. Therefore, the probed magnon wave vectors are also oriented in the same direction. Light inelastically scattered by magnons propagates along the backward path and is directed using a beam splitter (not shown) to the Fabry-P\'{e}rot interferometer for frequency analysis and photon counting.}
\end{figure}

The experimental setup is shown schematically in \Fig{f:2}. 
To achieve a large amplitude of the pumping magnetic field and, thus, a high magnon density, microwave pumping is supplied using half-wavelength microstrip resonators with a quality factor of about 25. The samples were placed in the middle of the resonators in the antinode of the microwave magnetic field. Both 50\,{\textmu m}- and 100\,{\textmu m}-wide microstrip resonators, with resonance frequencies of $13.2\,\rm{GHz}$, $13.6\,\rm{GHz}$, and $14.4\,\rm{GHz}$ were utilized. The pumping was performed with 1\,{\textmu s} long pulses with the peak power of up to $40\,\rm{W}$. A repetition interval of more than 200\,{\textmu s} ensured that the spin system is brought into equilibrium and that the thermal stability of the sample is maintained from pulse to pulse. The bias magnetic field $\B H$ was oriented perpendicular to the longitudinal axis of the resonators in the plane of the samples.

A probing laser beam of $532\,\rm{nm}$ wavelength is focused onto a spot with a diameter of about 20\,{\textmu m} in the parametrically pumped area of the YIG films. 
By setting the beam incidence angle $\Theta_{k \parallel}$ in a plane perpendicular to the film surface and oriented along the $\B H$ bias field (see \Fig{f:2}), one can selectively detect magnons with wave vectors $\B k_\|$ \cite{Bozhko2020a}. 
Varying $\Theta_{k \parallel}$ from $0$ to $\pm 58^\circ$ allows us to detect magnons with wave vectors ranging from $0$ to $\pm 2\cdot 10^5 \, \rm{cm}^{-1}$ with a resolution of about $ 1.5\cdot 10^3 \, \rm{cm}^{-1}$ \cite{Frey2021}.

In the backward Brillouin scattering geometry used, the component of the wave vector of the probing light lying in the film plane is reversed due to interaction with magnons. The component perpendicular to the film plane reverses direction by elastic reflection \cite{Serga2012}. Thus, it is necessary to ensure efficient and spatially homogeneous reflection of the probing beam after it passes through the film. This is achieved by covering the surface of the 6.7\,{\textmu m}-thick films facing the pump resonator with a thin dielectric mirror coating ($<1$\,{\textmu m}) \cite{Bozhko2020a, Frey2021, Kreil2021}. In the experiments with 5.6\,{\textmu m}-thick YIG films, light reflection occurred from the surface of the microstrip pumping resonator, which was in direct contact with the YIG film \cite{Serga2012, Bozhko2019}. 
The scattered light is sent to the Fabry-P\'{e}rot interferometer to analyze the Stokes and anti-Stokes spectral components, whose frequency shifts are equal to the magnon frequencies and whose intensities are proportional to the corresponding magnon densities. 

A temporal analysis of the magnon dynamics with a resolution of up to 400\,ps is achieved by recording the moments of detection of the scattered photons relative to the moment of application of the pump pulse \cite{Buttner2000, Frey2021}. The spatial analysis is realized by moving the sample together with the pump resonator relative to the BLS measurement point \cite{Buttner2000, Frey2021}.

The automation system thaTEC:OS (THATec Innovation GmbH) \cite{THATec} was used to control the experimental setup and to collect data.

\subsection{\label{ss:space} Spatial distribution of bottom magnons at various magnetic fields} 

Before proceeding to the experimental verification of the obtained theoretical results, we need to clearly define the region of experimental parameters at which one of the two mechanisms of the transition of parametric magnons to the bottom of the frequency spectrum prevails: the Kolmogorov-Zakharov cascade described in \Sec{s:FE-SP} and the kinetic instability of bottom magnons analyzed in \Sec{s:KI}. For this purpose, we  started with studying  the spatial distribution of the bottom magnons in the pumping area at different magnetic fields $H$. 

In these measurements, we employed a 100\,{\textmu m}-wide microstrip pumping resonator and a 6.7\,{\textmu m}-thick YIG film. The dielectric mirror coating of this sample allowed us to make measurements not only directly above the resonator but also in the surrounding regions of the YIG film. Owing to this mirror coating, the intensity of the back-scattered light is independent of the reflectivity of the microstrip material and its dielectric substrate and thus reflects the magnon density distribution well. The pumping frequency, determined by the length of the microstrip resonator, was 14.4\,GHz. The maximum available pumping power in our experiment was applied, which was 40\,W.

The incident angle $\Theta_{k \parallel}$ was set to $11^\circ$, which corresponded to the detection of magnons with wave numbers around $4.5\cdot 10^4 \, \rm{cm}^{-1}$. Therefore, only magnons from the bottom part of the spectrum (see Fig.\,\ref{f:real_spectrum}) were registered. The spatial distribution of the BLS intensity plotted in \Fig{f:exp_spatial} consists of the integrated anti-Stokes parts of two different spectra of inelastically scattered light measured for angles $\Theta_{k \parallel} = \pm 11^\circ$. Thus, the $\pm k$ bottom magnons are shown simultaneously in this figure.

At the longitudinal axis of the resonator, where the microwave pumping magnetic field is parallel to the bias field $H$, the conditions for the parallel pumping \cite{Neumann2009, Dzyapko2009} are realized. At the edges of the resonator, where the pump field is perpendicular to the field $H$, perpendicular pumping occurs \cite{Neumann2009, Dzyapko2009}. In the latter case, the direct pumping source is not the external electromagnetic field itself but the dynamic magnetization non-resonantly driven by this field. 

At a field $H=H_\mathrm{cr}$, corresponding to the minimal threshold of parametric instability, a blue region of low concentration of near-bottom magnons is well visible above the resonator. In this case, under the conditions of parallel pumping, parametric magnons with relatively small wave vectors $ \B k_\perp \perp \bm{H} $ and a frequency close to the ferromagnetic resonance frequency are excited, as schematically shown in \Fig{f:real_spectrum} by the two magenta squares. For this magnon group, the process of kinetic instability is forbidden by the laws of conservation of energy and momentum \cite{Melkov1991, Kreil2018}. The BLS intensity increases for lower and higher magnetic fields since the kinetic instability process allowed here leads to a higher magnon density at the bottom of the spectrum.

\begin{figure}[t]
\includegraphics[width=1\columnwidth]{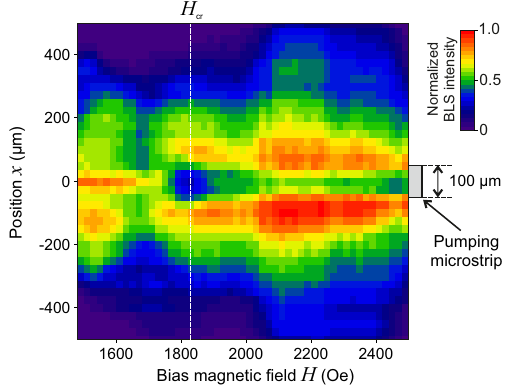}  
   \caption{\label{f:exp_spatial} Spatial distribution of the bottom magnon density across a 100\,{\textmu m}-wide microstrip pumping resonator measured for different magnetizing fields $H$. The pumping frequency is 14.4\,GHz.}
\end{figure}

\begin{figure*}[t]
 \includegraphics[width=1\textwidth]{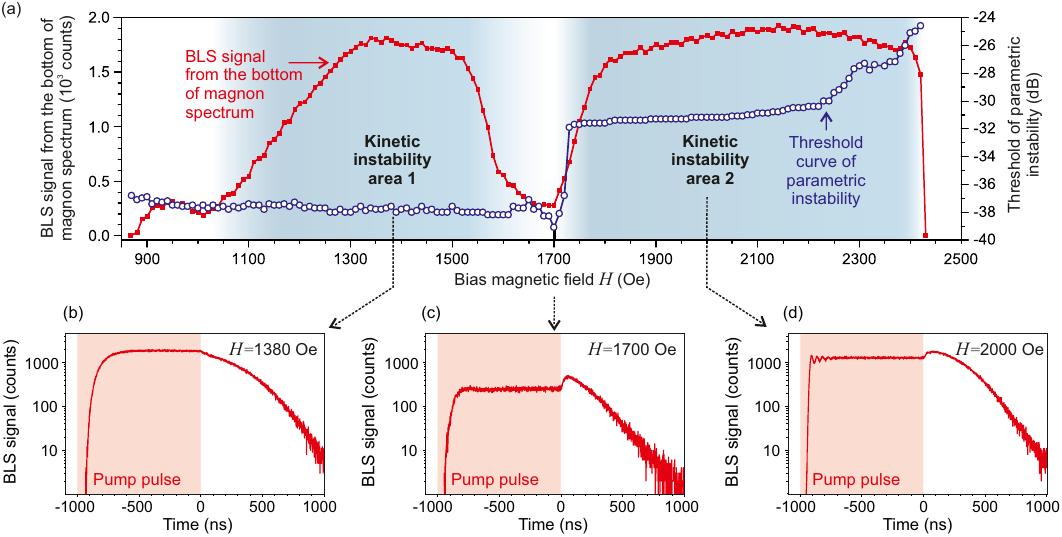}
  \caption{\label{f:KI} Panel (a): Measured parametric instability threshold (red squares) compared to the BLS signal from the bottom of the magnon spectrum (blue empty circles). Areas with blue shading, in which KI processes are allowed by the momentum and energy conservation laws, are marked as the ``Kinetic instability area 1 and 2''.
  Panels (b), (c), and (d) show the temporal dynamics of BLS signals from bottom magnons in the presence [panels (b) and (d)] and absence [panel (c)] of the kinetic instability process.
  The pumping frequency is 13.6\,GHz.  
}
\end{figure*}

At the same time, for perpendicular pumping, when magnons with large wave vectors directed at an angle of 45--55$^\circ$ to $\B H$ are excited \cite{Neumann2009}, there is no such strict prohibition. Consequently, no such significant difference in the bottom magnon density is observed at the sides of the microstrip resonator over the entire range of the bias magnetic fields.  
Moreover, the magnons excited by perpendicular parametric pumping propagate at non-zero angles to the resonator axis over quite considerable distances, thus expanding the spatial region of the overpopulated magnon gas. As a result, the area populated by bottom magnons born from the kinetic instability process is also expanded.

The conducted measurements allow us to determine the regions of the film in which the physical mechanisms of injection, thermalization, and spectral transfer of magnons analyzed in the previous sections are realized in the best and simplest way. Based on our findings, we can distinguish regions with parallel and perpendicular pumping of parametric magnons. Going forward, we will concentrate on the most effective and well-researched case of parallel pumping. In doing so, we avoid the need to take into account the spatial transport of parametric magnons and the associated effective damping.

\subsection{\label{ss:towards}From kinetic instability to BEC }

In the previous section, we observed a distinct forbidden area, surrounded by two regions of the allowed kinetic instability at lower and higher magnetic fields. To delve deeper into this phenomenon and to reveal the connection between the processes of kinetic instability and Bose-Einstein condensation, we have conducted an extensive experimental analysis of its properties under parallel parametric pumping.

To increase the amplitude of the pumping magnetic field, we used a resonator of 50\,{\textmu m} width. The BLS measurements were carried out in the 5.6\,{\textmu m}-thick film at a point on the longitudinal axis of this resonator, i.e., under the exclusive action of parallel pumping. The pumping frequency of 13.6\,GHz was determined by the resonator geometry. 
The obtained results are presented in \Fig{f:KI}.

An analysis of the conservation laws in thin YIG films performed in Ref.\,\cite{Kreil2018} for the same experimental conditions as in the current work shows that the kinetic instability is allowed in two ranges of magnetic field: Area 1 -- from $ H_{1,\rm min}\approx 1100\,$Oe to $ H_{1,\rm max}\approx1600\,$Oe; and Area 2 -- from  $H_{2,\rm min}\approx 1750\,$Oe to $ H_{2,\rm max}\approx2400\,$Oe. 

In Figure \Fig{f:KI}(a), the blue empty circles show the dependence of the parametric instability threshold $h\sb{th}$, introduced by \Eq{S-statE}, on the magnetic field. We see that in area\,1, below $H_\mathrm{cr}=1700$\,Oe, the value of $h\sb{th}$ is practically independent of $H$. Bearing in mind that with good accuracy $\omega_0 \propto H$ and that the wave vectors of the parametric magnons $\B k\sb{par}$ satisfy the relation $\omega_{\B k\sb{par}}=\omega\sb{pump}/2$ decreasing to zero when $H$ approaching $H_\mathrm{cr}$, we conclude that in this area the damping of parametric magnons $\gamma_{\B k}\propto h\sb{th}$ is practically independent of $\B k$. 
For many reasons  not discussed here, the relationship between  $\gamma_{\B k}$ and $h\sb{th}$ in area 2 is more complicated\,\cite{Neumann2009}, and here we leave the question about $k$-dependence of $\gamma_{\B k}$ for $H > H_\mathrm{cr}$ open. 

In the same \Fig{f:KI}(a), we also show by red squares the magnetic field dependence of the BLS signal intensity proportional to the total number of bottom magnons. It can be seen that in areas free from the kinetic instability, the number of the BLS counts is about 250, while with the KI active it jumps up to about 1800. 
As seen from the comparison with \Fig{f:exp_spatial}, this dependence correlates well with the density of bottom magnons on the magnetic field measured on the longitudinal axis of a wide (100\,\textmu m) pumping resonator. The difference in the value of the critical field $H_\mathrm{cr}$ in \Fig{f:exp_spatial} and \Fig{f:KI}(a) is due to the difference in the frequencies of parametric pumping.
We interpret these observations as evidence that the dominant contribution to the bottom magnons (above 80\%) comes from the kinetic instability and only a small part (below 20\%) originates from the cascade processes.}

The temporal dynamics of BLS signals from the bottom magnons are shown in the presence [see \Figs{f:KI}(b,d)] and absence [see \Figs{f:KI}(c)] of the kinetic instability process. 
In the latter case, one finds a significantly lower density of the bottom magnons during the pumping action and a jump-like increase in their density after turning off. This jump is the result of the effective population of the lowest energy states by the Kolmogorov-Zakharov scattering cascade after the pumping field is turned off and the disappearance of frequency-localized dense groups of parametric magnons. 

A similar but smaller jump in the magnon density can be seen in panel (d). In this case, the parametric magnons are excited closer to the bottom of the spectrum, and the Kolmogorov-Zakharov cascade plays a role comparable to the kinetic instability process. 

Concluding, we have to stress that when we compare our theoretical results with experimental findings, we must keep in mind that the pure impact of either the Kolmogorov-Zakharov cascade or the kinetic instability on particle transfer down to the BEC region is not fully realized. Instead, what we typically observe is a combination of these two mechanisms.
 
\subsection{\label{ss:prel}Frequency--wave-number distribution of magnons}  

In the sections\,\ref{ss:space} and \ref{ss:towards}, we presented qualitative arguments in favor of the important role of the kinetic instability process in transporting parametrically injected magnons to the lower end of their frequency spectrum. In this section, we share our experimental data on the frequency--wave-vector magnon distribution under microwave pumping. The analysis of this distribution reveals several specific nonlinear processes, including the four-wave scattering process \eqref{cons1} responsible for the kinetic instability. This gives us greater confidence that we are dealing with the kinetic instability phenomenon in our experiments, allowing us to compare it with our theory given in \Sec{s:KI}.

\begin{figure}[t] 
 \includegraphics[width=1\columnwidth]{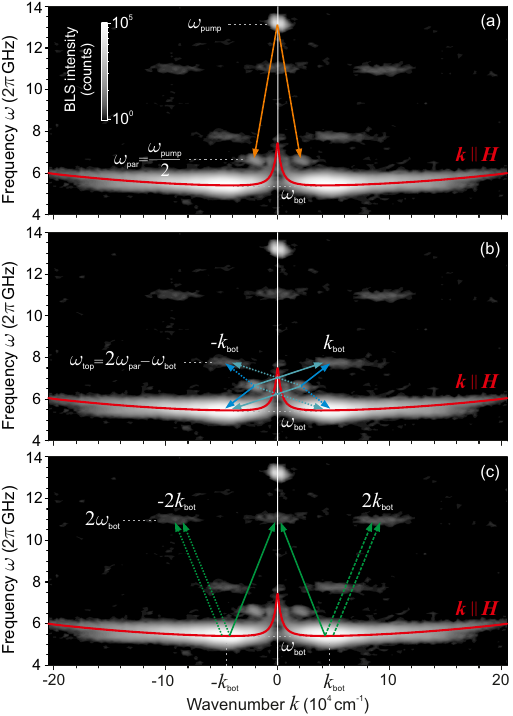}  
  \caption{\label{f:Experimental_spectrum_f-k} Frequency- and wave-vector-resolved BLS intensity spectrum of real and virtual magnons. The spectra were measured during the action of 13.2\,GHz microwave pumping on a YIG film magnetized in plane by the field $H=1885$\,Oe. The BLS intensity is proportional to the magnon density. The experimental intensity spectrum is shown together with the calculated magnon dispersion curve and diagrams showing the relevant quasiparticle scattering processes in the system: 
    (a) Orange arrows show the process of parametric pumping. The signal of virtual ``pump'' magnons is visible at the pumping frequency $\omega\sb{pump}$.   
    (b) The four-magnon kinetic instability processes leading to the appearance of virtual ``top'' magnons at $\omega\sb{top}$ frequency are shown by pairs of solid and dashed blue and cyan arrows.  
    (c) Virtual ``double-bottom'' magnons at the frequency $2\omega\sb{bot}$ arise due to the confluence of the bottom magnons at $\omega\sb{bot}$. Upward green arrows show the relevant confluence processes. }
\end{figure}

Figure\,\ref{f:Experimental_spectrum_f-k} shows the BLS intensity spectra $I(\omega, k_\|)$ of magnons with wave vectors $\B k = \B k_\| \parallel \bm{H}$. $I(\omega, k_\|)$ is proportional to the density of the corresponding magnons $n(\omega, k_\|)$.
The spectra $I(\omega, k_\|)$ were measured during microwave pumping at 13.2\,GHz in a YIG film magnetized in plane by the field $H=1885$\,Oe. 
The solid red line shows the calculated magnon frequency spectrum $\omega_{ k_\|}$, which has two minima $\omega\sb{bot}\=\min_{k_\|}\{\omega_{k_\|}\}\approx 4\cdot (2\pi)$\,GHz at $k_{\|}=\pm k\sb{bot}$ with $k\sb{bot}\approx 4\cdot 10^3\,$cm$^{-1}$. The two brightest spots in the vicinity of the bottom of the magnon spectra $\omega\sb{bot}$, $\pm k\sb{bot}$ originate from the ``bottom'' magnons associated with the left and right BEC states. They are spread around the bottom of the spectrum due to the scattering of the bottom magnons on the parametric ones $\omega(\B k\sb{bot}+\B \kappa_1)+\omega (\B k\sb{par}+\B \kappa_2 )= \omega(\B k\sb{bot}+\B \kappa_3)+\omega( \B k\sb{par}+ \B \kappa_4 )$, which was discussed in \Sec{s:KI}. Here,  $\B \kappa_1+\B \kappa_2=\B \kappa_3+\B \kappa_4$ and $\kappa_j\ll k\sb{bot}$. 

Above these brightest spots we see three spots with $\omega\approx 2\omega\sb{bot}$ and $k\sb{left}\approx -2 k\sb{bot}$, $k\sb{center}\approx 0$, and $k\sb{right}\approx 2 k\sb{bot}$. They are related to the confluence of two bottom magnons, as shown by green arrows in \Fig{f:Experimental_spectrum_f-k}(a):

i) left spot, $ \omega_{-k_{\|}}+ \omega_{-k_{\|}} \Rightarrow 2 \omega\sb{bot}$ and $k=-2 k\sb{bot}$; 
 
ii) central spot, $\omega_{-k_{\|}}+ \omega_{+k_{\|}} \Rightarrow 2 \omega\sb{bot}$ and $k=0$; 

iii) right spot, $\omega_{+k_{\|}}+ \omega_{+k_{\|}} \Rightarrow 2\omega\sb{bot}$ and $k = 2 k\sb{bot}$.\\
Note that  neither $\omega=2\omega\sb{bot}$ with $k=\pm 2 k\sb{bot}$ nor $\omega=2\omega\sb  {bot}$ with $k=0$ are eigenmodes of the YIG film in our magnetization geometry. Therefore, what we see are the off-resonant waves driven by an appropriate nonlinearity, i.e., virtual magnons, called ``double-bottom virtual magnons'' in \cite{Lvov2023}.  

In \Fig{f:Experimental_spectrum_f-k}(a), the BLS spectra $I(\omega, k_\|)$ are supplemented by two down-pointing orange arrows showing the process of parametric pumping by the external quasi-homogeneous microwave field with wave vector $\B k\sb {pump}\approx 0$ and frequency $\omega\sb{pump}$. Precisely at this position, we see a rather bright spot, indicating virtual ``pumped'' magnons~\cite{Lvov2023}. 
At the same time, at the frequency of the parametrically pumped real magnons $\omega\sb{par}=\omega\sb{pump}/2$ we see no BLS response because the wave numbers of the parametric magnons are pretty large and lie outside the sensitivity range of our BLS setup.

Two more spots visible at $\omega\sb{top}=2\omega\sb{par}-\omega\sb{bot}$ and $k\sb{top}=\pm k\sb{bot}$ indicate magnons with frequency of the top magnons involved in the kinetic instability process [see \Eq{cons1}].

If so, then the top magnons must have wave vectors $\B k\sb{top}=\B k_1+\B k_2\mp \B k\sb{bot}$, where $\B k_1$ and $\B k_2$ are the wave vectors of the parametric magnons with $\omega(\B k_1)=\omega(\B k_2)=\omega\sb{pump}/2$. 
Assuming for a rough estimate that $\omega\sb{top} > \omega\sb{bot}$, we conclude that $k \sb{top} \gtrsim k_1$ and $k \sb{top} \gtrsim k_2$, meaning that the top magnons lie outside the sensitivity region of our BLS setup, i.e., they are invisible in \Fig{f:Experimental_spectrum_f-k}. 
The origin of the two spots in \Fig{f:Experimental_spectrum_f-k}(b) at frequency $\omega\sb{top}=\omega\sb{pump}-\omega\sb{bot}$, which is consistent with \Eq{cons1}, but with wave vectors $\B k\sb{top}=\pm \B k\sb{bot}$ was clarified in \cite{Lvov2023}. 
It was stressed that the theory of kinetic instability is formulated in the framework of the weak-wave kinetic equation, which assumes weak correlations of the wave phases. As a result, the scattering\,\eqref{scat} of real magnons has a stochastic nature and appears only as the second-order perturbation of the four-wave interaction amplitudes $W_{\B 1,\B 2}^{\B 3, \B 4}$, \Eq{4-St}. 
Nevertheless, in our particular case with a large population of parametric and bottom magnons, there are strong, externally determined, phase correlations of the scattering waves. 
In particular, the full phase correlation in the pairs of parametric waves with  $\pm \B k\sb{par}$ arises due to their interaction with the space-homogeneous pumping field \cite{Lvov1993}. Given this correlation, 
$  | \< a_{\B k\sb{par}}a_{-\B k\sb{par}}
   \exp [i \omega \sb{par}] \>|= \< |a_{\B k\sb{par}}|^2 \> \=n_{\B k\sb{par}}$.
This allows us to consider a pair of parametric magnons $(a_{\B k\sb{par}}a_{-\B k\sb{par}})$ as a ``single'', coherent wave object with the frequency $2\omega\sb{par}=\omega\sb{pump}$ and phase being the sum of the phases of the waves composing the pair. Therefore, four-wave scattering \eqref{scat} with $\B k_1= -\B k_2 $, $\B k_3=\pm \B k\sb{bot} $, due to its dynamic nature, appears much stronger than stochastic scattering   with $\B k_1 \ne -\B k_2$, being now proportional to the first power of the interaction amplitude $W_{\B 1,\B 2}^{\B 3, \B 4}$, and producing the driving force \cite{Lvov2023}
\begin{equation}\label{F}
F=\sum_{\B k\sb{par}}T^{\B k\sb{par},-\B k\sb{par}}_{\pm \B k\sb{bot},\mp \B k\sb{bot}}a^*_{\B 2} a_{\B k\sb{par}}a_{-\B k\sb{par}} \ .
\end{equation}
This force has the same frequency \eqref{cons1} as that of real top magnons [see \Eq{condA}] but with the different wave vector $\bm{q}\sb{top} = \mp \B q\sb{bot}$. This force excites off-resonant magnons, seen in two bright spots, as discussed earlier.

\subsection{\label{ss:power}Pumping power dependence of the parametric and bottom magnon numbers}
In Fig.\,\ref{F:3}, we plot theoretical predictions for the dependence of the total number of parametric and bottom magnons, $N\sb{par}$ and $N\sb{bot}$ as a function of the relative amplitude of the microwave pumping field $\dfrac{h V}{\gamma\sb{par}} = \dfrac{h}{h\sb{th}}$. 
Our theory considers only the kinetic instability mechanism of the transfer of parametric magnons to the lower magnon region and does not take into account the mechanism of the step-by-step Kolmogorov-Zakharov cascade. 
Therefore, for the comparison of theory and experiment, we have to choose the range of bias magnetic fields $H$,  where kinetic instability is allowed, see Fig.\,\ref{f:KI}.  In the range of lower magnetic fields $H$, denoted ``kinetic instability area 1'', the wave numbers of the parametric magnons are large and cannot be detected by BLS spectroscopy. 
For this reason, we have chosen for the comparison the magnetic field range designated as ``kinetic instability area\,2'',  taking for the sake of concreteness $H=1885\,$Oe as in Fig.\,\ref{f:Experimental_spectrum_f-k}.

\begin{figure}[t] 
\includegraphics[width=1\columnwidth]{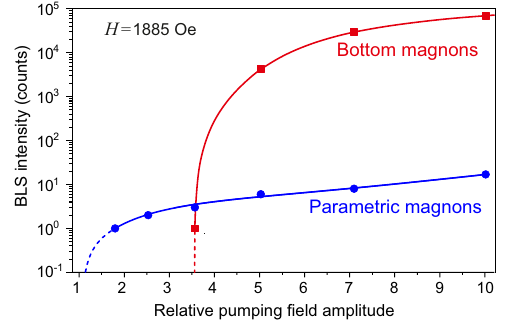}
\caption{\label{f:power} Dependence of the number of BLS counts from the parametric magnons $\C N\sb{par}$ (blue circles) and that from the bottom magnons $\C N\sb{bot}$ (red squares) on the pumping field amplitude $h$ normalized by the parametric instability threshold $h\sb{th}$. 
$H=1885$\,Oe. The pumping frequency is 13.2\,GHz. The solid red and blue curves represent an interpolation and an extrapolation of the corresponding experimental data using cubic splines. The dashed lines are extrapolations of the experimental dependencies in the region of relatively low levels of parametric pumping.}
\end{figure}

In Fig.\,\ref{f:power}, one can see the numbers of BLS counts $\C N\sb{par}$ and $\C N\sb{bot}$ obtained from the parametric and bottom magnons at different pumping supercriticalities $h/h\sb{th}$ and represented by blue circles and red squares, respectively. 
Assuming a smooth dependence of $\C N\sb{par}$ and $\C N\sb{bot}$ on $h/h\sb{th}$, we used the procedure of interpolating the experimental data by a cubic spline, which resulted in the blue and red solid lines.  
Moreover, we used the available data to extrapolate the desired dependence of $\C N\sb{par}$ and $\C N\sb{bot}$ on $h/h\sb{th}$ to the region of low pumping powers, where the low signal-to-noise ratio did not allow for experimental observations. These results are shown by blue and red dashed lines. By finding the value of $h$, at which $\C N\sb{par}\to 0$, we accurately estimated the threshold value $h\sb {th}$ for parametric instability. This value was used to normalize the scale of the abscissa axis in Fig.\,\ref{f:power}. 
Since the sensitivity of the BLS setup to parametric and bottom magnons is different, the ratio $\C N\sb{par}/\C N\sb{bot}$ does not reflect the ratio of their occupation numbers $N\sb{par}/ N\sb{bot}$. However, these experimental curves correctly reproduce the dependence of $N\sb{par}$ and $N\sb{bot}$ on $h$ in units of the threshold field of parametric instability.

\begin{figure*}[t]
\includegraphics[width=\textwidth]{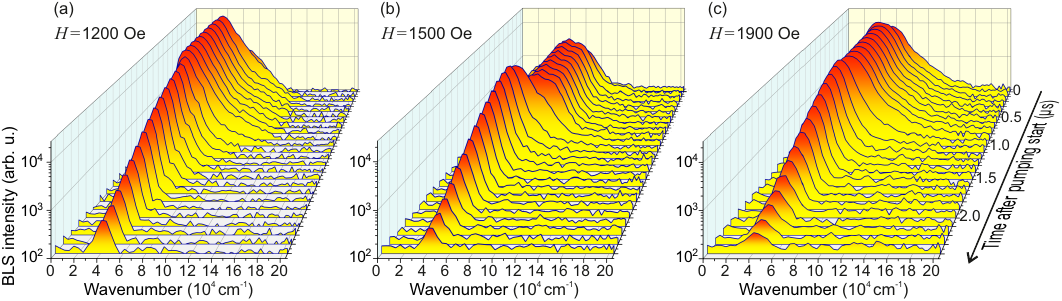}
\caption{\label{f:X1} 
Distributions of the bottom magnons by wave numbers at different time delays after pumping was turned on.
Panels (a), (b), and (c) show three-dimensional plots for three different bias magnetic fields $H$. 
The pumping pulse has a duration of 1\,{\textmu s} (note the direction of the time axis). To improve the signal-to-noise ratio, we integrated the spectra in the time window of 100\,ns and over the entire frequency range of the bottom magnons. 
The pumping frequency is 13.2\,GHz.
    	}  
\end{figure*}

A comparison of the theoretically predicted dependencies of $N\sb{par}$ and $N\sb{bot}$ on $h/h\sb{th}$ shown in Fig.\,\ref{F:3} with the experimental results shown in Fig.\,\ref{f:power} demonstrates a fairly good qualitative agreement. 
This is a strong argument for the validity of our nonlinear theory of kinetic instability, which evidences that this theory captures the essential physical mechanisms governing the phenomenon.  
 
At the same time, the experimental data shown in Fig.\,\ref{f:power} does not demonstrate a pronounced saturation of the dependence $N\sb{par}(h/h\sb{th})$, as predicted by the theory. This may be due to some secondary effects, such as the relatively small contribution of the Kolmogorov-Zakharov cascade to the particle flux towards lower frequencies, which is not yet considered in our theory.  Another possible origin of the discrepancy is a change in the magnon excitation region caused by a downward frequency shift of the magnon spectrum due to a decrease in magnetization at high pumping powers and a consequent increase in the efficiency of parametric pumping. 

\subsection{\label{ss:evol}Time evolution at different magnetic fields}
In Sections \ref{ss:space}, \ref{ss:towards}, and \ref{ss:prel}, we theoretically analyzed different aspects of the behavior of bottom magnons during parametric pumping. We concluded that the kinetic instability essentially contributes to the transfer of magnons to their frequency minimum. 
In addition, the nonlinear kinetic instability theory developed in Sect.\,\ref{ss:NLT} also accounts for the scattering of the bottom magnons on the parametric magnons and shows that this scattering leads to a broadening of the near-bottom magnon distribution in the vicinity of $\B k\sb {bot}$, $\omega\sb {bot}$. If this scattering ceases, we expect bottom magnons to evolve into BECs, narrowing their frequency and wave vector spectra.

Fortunately, the theory of kinetic instability suggests how to test this statement experimentally: one can study the evolution of the distribution of the near-bottom magnons over frequencies and wave vectors after turning off the parametric pumping. The frequency of parametric magnons is higher than that of the bottom magnons. Therefore, we can expect that their relaxation $\gamma\sb{par}$ in the linear regime is larger than the linear relaxation $\gamma\sb{bot}$ of the bottom magnons. When the number of parametric magnons is very large, the kinetic instability opens a very efficient additional dissipation channel for parametric magnons. This means that even if, in the linear regime, the relaxation rates of parametric and bottom magnons are approximately equal, in the nonlinear regime, the relaxation rate of parametric magnons is much larger than that of the bottom magnons. Consequently, after pumping is turned off, there is a period during which parametric magnons are practically absent, while bottom magnons continue to exist, and the expected narrowing of their spectrum can be detected.

According to Ref.\,\cite{Noack2021}, the distribution of magnons in a system has been significantly narrowed down, as measured by detecting electromagnetic radiation in the frequency domain. The resolution of BLS spectroscopy is not fine enough to record this effect, but we have a good resolution in the wave-number domain, as shown in \Fig{f:Experimental_spectrum_f-k}. By integrating the frequency distribution of the bottom magnons, in \Fig{f:X1} we plotted the BLS intensity versus the magnon wave number $k_\|$ and time for different magnetic fields after turning on the parametric pumping.

Examining the data, we can see that during the first 1\,{\textmu s} of the pumping pulse, the peak of the distribution of bottom magnons in $\bm{k}$-space has a relatively constant and broad shape. However, after the pumping is switched off, this peak quickly narrows, which aligns with our theoretical expectations. Eventually, the peak width reaches the limit of the wave number resolution.

Figure\,\ref{f:X1}(a) with $H=1200\,$Oe corresponds to region 1 of the kinetic instability, see \Fig{f:KI}, while \Fig{f:X1}(c) with $H=1900\,$Oe corresponds to region 2 of the kinetic instability. In these cases, the parametric magnons are transferred directly to the bottom of the spectrum.  

In \Fig{f:X1}(b), where $H=1500\,$Oe, the Kolmogorov-Zakharov cascade plays an important role in the magnon distribution process. This results in a significant portion of magnons being distributed between the parametric and the bottom parts of the $\bm{k}$-space. When the pumping is switched off, these magnons continue to move towards the bottom, creating an intense hump that is clearly visible in \Fig{f:KI}(b).

\begin{figure} 
 \includegraphics[width=.86\columnwidth]{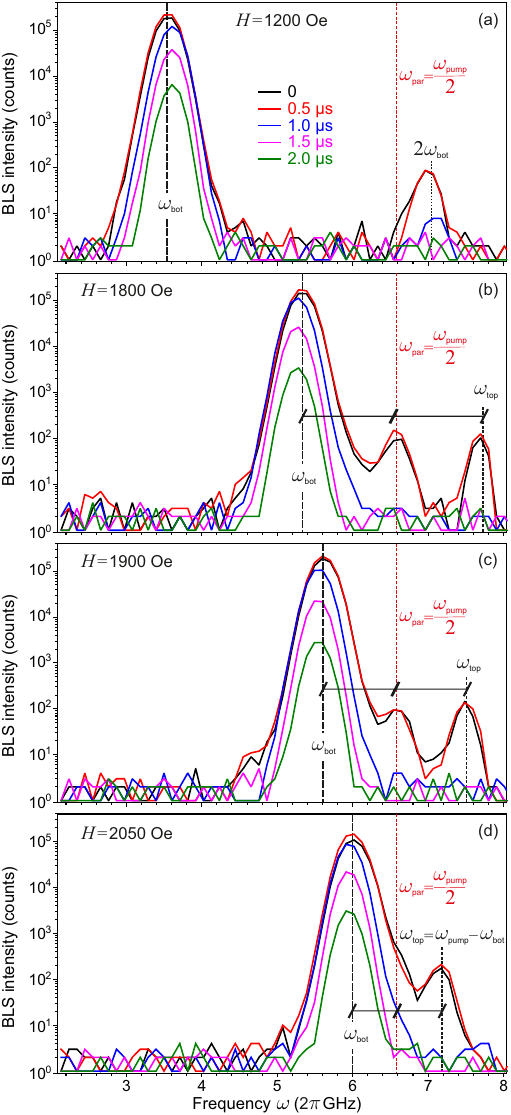}
\caption{
		\label{f:Spectra_vs_Time} 
		Frequency spectra at varios bias magnetic fields and varios times after pumping is switched off. The pumping frequency is 13.2\,GHz.
	    }    
\end{figure}
Another perspective on the BLS spectra is shown in \Fig{f:Spectra_vs_Time}.
We now integrated them over wave number, obtaining frequency spectra for different magnetic fields and times after switching off the parametric pumping.  {In panel (a) we plot the results for the small magnetic field $H=1200\,$Oe. The value of $\omega\sb{bot}/(2\pi)\approx 3.5$\,GHz is shown as a vertical black dashed line, while $2\omega\sb{bot}/(2\pi)$ is shown as a vertical black dotted line. One sees an intense peak of the bottom magnons and a much smaller peak (by about three orders of magnitude) of the double-bottom virtual magnons. The parametric magnons with the frequency $\omega\sb{par}=\omega\sb{pump}/2$, shown by the vertical red dotted line, are not seen in this panel. They have large wave numbers that are outside of the accesible zone for our BLS setting. For a larger magnetic field, the wave numbers decrease and we see peaks of the parametric magnons in \Figs{f:Spectra_vs_Time}(b,c). For the largest magnetic field $H=2050\,$Oe, shown in \Fig{f:Spectra_vs_Time}(d), the frequency of the parametric magnons is very close to the very intense peak of the bottom magnons. Therefore the peak of the parametric magnons just slightly disturbs the peak of the bottom magnons. 

In all \Figs{f:Spectra_vs_Time}(b)-\ref{f:Spectra_vs_Time}(d) for $H>1200\,$Oe, the frequency $2\omega\sb{bot}$ is outside the accesible zone and the peak $2\omega\sb{bot}$ seen in \Fig{f:Spectra_vs_Time}(a) has disappeared. Instead, we see a peak at $\omega\sb{top}$ that is exactly at the required position $\omega\sb{pump}-\omega\sb{bot}$. 
This is another confirmation that the kinetic instability essentially contributes to the population of the bottom magnons. 
 
As an additional support for the kinetic instability picture, we note the absence of a continuous magnon distribution between $\omega\sb{par}$ and $\omega\sb{bot}$, expected in the case of the Kolmogorov-Zakharov step-by-step cascade.

In summary, the experimental findings and discussions presented in this section lead us to conclude that the primary cause of the transfer of magnons from the region of their parametric pumping with the frequency of $\omega\sb{par}=\omega\sb{pump}/2$ to the bottom of their frequency spectrum $\omega\sb{bot}$ is the kinetic instability discussed in  \Sec{s:KI}. The experiments confirm that the main mechanism that limits the number of bottom magnons is their feedback effect on the parametric magnons, as described in Section \Sec{ss:S-theory}. Additionally, the experiments confirm that the scattering of the bottom magnons on parametric ones, described in \Sec{ss:NLT}, plays the leading role in widening the bottom magnons' distributions.

%%%%%%%%%%%%%%%%%%%%%%%%%%%%%%%%%
  
\section{\label{s:summary}Summary}
 We presented a systematic and comprehensive description of the physical mechanisms leading to the  Bose-Einstein condensation of quasiparticles. Unlike the atomic BEC forming in the thermodynamic equilibrium conditions, the quasiparticles condense under conditions of flux equilibrium and represent 
a nonlinear wave system with energy pumping and dissipation. We find the conditions under which Bose-Einstein condensation of quasiparticles is possible. The first and obvious constraint that we took into account is the conservation (or almost complete conservation) of the total number of quasiparticles in the non-linear processes. This means that the four-wave scattering processes $2\Leftrightarrow 2$ must dominate over the three-wave processes near the bottom of the frequency spectrum. 

We started in \Sec{s:FE-WP}, with the pumping weak enough to keep the wave system close to the thermodynamic equilibrium. In this case, it is necessary to simply balance the pumping and damping rates of the total number of quasiparticles $N\sb{tot}$ and the total energy in the system, giving the conditions under which the total number $N\sb{tot}$ of quasiparticles in the system exceeds the number of quasiparticles $N\sb{gas}$. The excess $N\Sb{BEC}=N\sb{tot}-N\sb{gas}$ can occupy excited energy levels and create a BE-condensate at the zero energy level. 

The situation with strong pumping is less straightforward. It is necessary to consider the kinetic wave equation to describe the transport of quasiparticles from the pumping range to the lower part of the wave frequency spectrum. In \Sec{s:FE-SP}, we have done this under the assumption of the scale invariance of the system. In this case, analytic solutions of the kinetic equation are available. For the $2 \Leftrightarrow 2$ scattering, the kinetic equation can have two differently oriented solutions for the energy and particle fluxes. We have specified the conditions under which the particle flux is oriented toward small $k$, allowing the creation of a BE-condensate. 

An even more complicated scenario is realized by a super-strong injection of quasiparticles into a narrow frequency range, for example, by high-power parametric pumping. In this case, the relaxation rate of quasiparticles becomes negative, first at small wave vectors $k$ at the lower part of the frequency spectrum. This leads to the phenomenon of exponential growth of the number of quasiparticles with small $k$, known as kinetic instability. In \Sec{s:KI}, we developed a nonlinear theory of kinetic instability that considers the feedback of unstable bottom quasiparticles on their source -- the parametrically excited quasiparticles. This theory also accounts for the $2\Leftrightarrow 2$ scattering of bottom quasiparticles on parametric quasiparticles, which broadens the bottom quasiparticle packet. 

In the last section \Sec{s:EXP}, we presented an experimental study of BE magnon condensation in yttrium iron garnet thin films using Brillouin light scattering spectroscopy. The theoretical and experimental results are in qualitative agreement. Therefore, we conclude that, if allowed by conservation laws, the kinetic instability serves as the dominant source of bottom magnons in the vicinity of their BE-condensation points, and that the nonlinear theory of kinetic instability developed in \ref{s:KI} describes the main physical mechanisms of this process quite well. The above comparison of our analytical findings and  experimental observations opens new directions for further studies of this phenomenon.

%%%%%%%%%%%%%%%%%%%%%%%%%%%%%%%%%
\section*{Acknowledgements} 
This study was funded by the Deutsche Forschungsgemeinschaft (DFG, German Research Foundation) -- TRR 173 -- 268565370 Spin+X (Project B04). D.A.B. acknowledges support by grant ECCS-2138236 from the National Science Foundation of the United States. V.S.L. was partly supported by NSF-BSF grant \# 2020765. 
 
%\bibliography{BECinNice.bib,2021-07-Lvov-Refs.bib,2021-07-BEC-general.bib} 
\bibliography{BECinNice.bib} 

\end{document}